\documentclass[11pt, a4paper]{article} 
\usepackage{slashed,jheppub,multirow,relsize,soul}
\usepackage[normalem]{ulem}
\usepackage{color}
\usepackage{subfig}
\usepackage[capitalise]{cleveref}
\usepackage{amsmath}
\usepackage{booktabs}
\usepackage{appendix}
\usepackage{tabularx}
\usepackage{changepage} 
\makeatletter
\def\@fpheader{\relax}
\makeatother

\def\ie{\emph{i.e. }}

\def\th#1#2{\ensuremath{\theta_{#1#2}}}
\def\s#1#2{\ensuremath{s_{#1#2}}}
\def\c#1#2{\ensuremath{c_{#1#2}}}
\def\Dm#1#2{\ensuremath{\Delta m^2_{#1#2}}}

\newcommand{\sublabel}[1]{\subfloat{\label{#1}}}

\title{Precision neutrino experiments vs the Littlest Seesaw}

\author[a]{Peter Ballett,}
\author[b]{Stephen F. King,}
\author[a]{Silvia Pascoli,}
\author[b,c]{Nick W. Prouse}
\author[a]{and TseChun Wang}

\affiliation[a]{Institute for Particle Physics Phenomenology, Department of
Physics, Durham University, South Road, Durham DH1 3LE, United Kingdom.}
\affiliation[b]{School of Physics and Astronomy, University of Southampton, SO17 1BJ Southampton, United Kingdom.} 
\affiliation[c]{Particle Physics Research Centre, School of Physics and Astronomy, Queen Mary University of London, Mile End Road, London E1 4NS, United Kingdom.} 

\emailAdd{peter.ballett@durham.ac.uk}
\emailAdd{king@soton.ac.uk}
\emailAdd{silvia.pascoli@durham.ac.uk}
\emailAdd{n.prouse@soton.ac.uk}
\emailAdd{tse-chun.wang@durham.ac.uk}

\preprint{IPPP/16/113}

\abstract{
We study to what extent upcoming precision neutrino oscillation
experiments will be able to exclude one of the most predictive models of
neutrino mass and mixing: the Littlest Seesaw. We show that this model provides
a good fit to current data, predicting eight observables from two input
parameters, and provide new assessments of its predictions and their
correlations. 
We then assess the ability to exclude this model using simulations of upcoming
neutrino oscillation experiments including the medium-distance reactor
experiments JUNO and RENO-50 and the long-baseline accelerator experiments DUNE
and T2HK.
We find that an accurate determination of the currently least well measured
parameters, namely the atmospheric and solar angles and the CP phase $\delta$,
provide crucial independent tests of the model. 
For $\theta_{13}$ and the two mass-squared differences, however, the model's
exclusion requires a combination of measurements coming from a varied experimental programme.
Our results show that the synergy and complementarity of future experiments will play a vital role in efficiently discriminating between predictive models of neutrino flavour, and hence, towards advancing our understanding of neutrino oscillations in the context of the flavour puzzle of the Standard Model.
}

\begin{document} 

\maketitle

\section{Introduction}
The framework of neutrino masses and mixing for explaining
neutrino oscillations --- the first direct experimental
evidence for physics beyond the Standard Model ---
is now firmly established~\cite{nobel}. All three mixing angles together with
the size of the two mass-squared differences have been measured, with
experimental efforts now focused on determining the final few
unknowns: the ordering and scale of the neutrino masses; the value of the Dirac
phase $\delta$; and a precision measurement of the angle \th23 including, if
non-maximal, its octant. Although there is some as yet
inconclusive evidence for $\delta$ in the third or fourth quadrant, as well as
for normal ordering (NO) and non-maximal atmospheric mixing, we
rely on the next generation of oscillation experiments to set these issues to
rest.

On the theoretical side, however, the origin of neutrino masses and mixing
remains unknown with many possible models considered viable (for
reviews see e.g.~\cite{King:2013eh,King:2015aea}). A large proportion of these
models are based on the classic seesaw mechanism, involving heavy right-handed
Majorana neutrinos~\cite{seesaw}, providing both a mechanism for generating the
neutrino masses and a natural explanation for their smallness. However, in
order to make predictions that can be probed experimentally, seesaw models
require additional assumptions or constraints~\cite{King:2015sfk}.

To accommodate the three distinct light neutrino masses which
drive the oscillation phenomenon, the seesaw mechanism requires at least two
right-handed neutrinos~\cite{King:1999mb}. In order to reduce the number of
free parameters still further to the smallest number possible, and hence
increase predictivity, various approaches to the two right-handed neutrino
seesaw model have been suggested\footnote{In seesaw models with two right-handed neutrinos, including those discussed in this paper, a hierarchical spectrum of left-handed neutrino masses is obtained where the lightest left-handed neutrino is massless.}, such as postulating one~\cite{King:2002nf} or
two~\cite{Frampton:2002qc} texture zeroes in the Dirac mass matrix in the
flavour basis (i.e. the basis of diagonal charged lepton and right-handed
neutrino masses). However, such two texture zero models are now phenomenologically
excluded~\cite{Harigaya:2012bw} for the case of a normal neutrino mass
hierarchy.
The minimal two right-handed neutrino model with normal hierarchy
which can accommodate the known data of neutrino mixing involves
a Dirac mass matrix with one texture zero and a characteristic form known as
the Littlest Seesaw model~\cite{King:2013iva}. The Littlest Seesaw model may be
embedded in unified models of quarks and leptons in~\cite{Bjorkeroth:2015ora}.
It leads to successful leptogenesis where the sign of baryon asymmetry is
determined by the ordering of the heavy right-handed neutrinos, and the only
seesaw phase $\eta$ is identified as the leptogenesis phase, linking
violation of charge parity symmetry (CP) in the laboratory with that in the early universe
\cite{Bjorkeroth:2015tsa}.

The Littlest Seesaw model can be understood as an example of
sequential dominance (SD)~\cite{King:1998jw} in which one right-handed neutrino
provides the dominant contribution to the atmospheric neutrino mass\footnote{With the lightest neutrino massless, $m_1=0$, we refer to the two non-zero masses as the \textit{solar neutrino mass} and the \textit{atmospheric neutrino mass}, corresponding to the square roots of the experimentally measured solar and atmospheric neutrino mass splittings $m_2=\sqrt{\Dm21}$ and $m_3=\sqrt{\Dm31}$ respectively.}, leading to approximately maximal atmospheric mixing, while the other right-handed neutrino
gives the solar neutrino mass and controls the solar and reactor mixing
as well as the magnitude of CP violating effects via $\delta$.
SD generally leads to normal ordering and a reactor angle which is bounded by
$\theta_{13}\lesssim m_2/m_3$~\cite{King:2002nf}, proposed a decade before the
reactor angle was measured~\cite{nobel}. Precise predictions for the reactor
(and solar) angles result from applying further constraints to the
Dirac mass matrix, an approach known as constrained sequential dominance
(CSD)~\cite{King:2005bj}.  For example, keeping the first column of the Dirac
mass matrix proportional to $(0,1,1)^T$, a class of CSD($n$) models has
emerged~\cite{King:2005bj, Antusch:2011ic, King:2013iva, King:2013xba,
Bjorkeroth:2014vha} corresponding to the second column proportional to
$(1,n,(n-2))^T$, with a reactor angle approximately given
by~\cite{King:2015dvf} $\theta_{13} \sim (n-1) \frac{\sqrt{2}}{3}
\frac{m_2}{m_3}$.  The Littlest Seesaw model corresponds to $n=3$ with a fixed
seesaw phase $\eta = 2\pi/3$.

It was recently realised that the alternative form of the Littlest Seesaw model
with second column $(1,1,3)^T$ and seesaw phase $\eta = -2\pi/3$ (also proposed
in~\cite{King:2013iva}) may be enforced by an $S_4\times U(1)$ symmetry,
putting this version of the Littlest Seesaw model on a firm theoretical
foundation~\cite{King:2016yvg} in which the required vacuum alignment emerges
from symmetry as a semi-direct model~\cite{King:2009ap}.  In general the
Littlest Seesaw model is an example of trimaximal~TM$_1$
mixing~\cite{Xing:2006ms,Albright:2008rp}, in which the first column of the
tri-bimaximal mixing matrix~\cite{Harrison:2002er} is preserved, similar to the
semi-direct model of trimaximal~TM$_1$ mixing that was developed
in~\cite{Luhn:2013vna}.  To fix the seesaw phase, one imposes a CP symmetry in
the original theory which is spontaneously broken, where,
unlike~\cite{Ding:2013hpa}, there is no residual CP symmetry in either the
charged lepton or neutrino sectors, but instead the phase~$\eta$ in the
neutrino mass matrix is fixed to be one of the cube roots of unity due to a
$Z_3$ family symmetry, using the mechanism proposed in~\cite{Antusch:2011sx}.

As explained in more detail later on, the Littlest Seesaw model
predicts all neutrino masses and mixing parameters in terms of two or three
parameters, and it has been shown that the model is in agreement with all
existing data, for a suitable range of its internal parameters
\cite{Bjorkeroth:2014vha}.  The model makes some key predictions
about the neutrino mass spectrum, that the lightest neutrino is massless
$m_1=0$ and that normal ordering obtains $\Dm31>0$, which offer a means to
exclude it via the observation of neutrinoless double beta decay, the
measurement of the beta-decay end-point, or from cosmological measurements, as
well as any measurement of NO from neutrino oscillation searches. However, it
also provides a rich set of predictions and correlations for the mixing angles
and phases. In this paper, we assume Normal Hierarchy ($m_1=0$ and NO), and
study how the future long- and medium-baseline oscillation programme will be
able to test this model through the precision measurement of the oscillation
parameters.
%

The layout of the paper is as follows:  in \cref{sec:csd} we define the
Littlest Seesaw models discussed above, and express some of the predictions in
terms of exact sum rules of the neutrino oscillation parameters.  In
\cref{sec:data} the Littlest Seesaw models are confronted with
existing oscillation data and we show the precise predictions
made once this data is taken into account.  \Cref{sec:sens} then covers how the
predictions of the models could be probed at future experimental facilities,
showing the sensitivities of experiments to exclude the models and the combined
measurements required to do so. We end with some concluding
remarks in \cref{sec:conc}.

\section{Littlest Seesaw models of neutrinos}\label{sec:csd}

Sequential dominance models of neutrinos arise from the proposal that, via the
type-I seesaw mechanism, a dominant heavy right-handed (RH) neutrino is mainly
responsible for the atmospheric neutrino mass, a heavier subdominant RH
neutrino for the solar neutrino mass, and a possible third largely decoupled RH
neutrino for the lightest neutrino mass~\cite{King:1998jw}. This leads to the
prediction of normal neutrino mass ordering and, in the minimal case containing
just the dominant and subdominant right-handed neutrinos, the lightest neutrino
must be massless. Constrained sequential dominance (CSD) constrains these
models further through the introduction of flavour symmetry, with the indirect
approach used to fix the mass matrix from vacuum alignments of flavon
fields~\cite{King:2005bj}. A family of such models, parameterized by $n$, either
integer or real using the flavour symmetry groups $S_4$ or $A_4$ respectively,
predicts the CSD($n$) mass matrix for left-handed
neutrinos~\cite{King:2013iva,King:2015dvf}. This model is also known as the
Littlest Seesaw (LS) model since it provides a physically viable seesaw model
with the fewest number of free parameters. After integrating out
the heavy neutrinos, the resulting left-handed light effective Majorana
neutrino mass matrix\footnote{We follow the Majorana mass Lagrangian
convention $-\frac{1}{2}\overline{\nu_L}m^\nu \nu^c_L$. } in the charged-lepton
flavour basis is given by
\begin{equation}
m^\nu=m_a
\begin{pmatrix}
0 & 0 & 0 \\
0 & 1 & 1 \\
0 & 1 & 1
\end{pmatrix}
+m_be^{i\eta}
\begin{pmatrix}
1 & n & (n-2)\\
n & n^2 & n(n-2)\\
(n-2) & n(n-2) & (n-2)^2
\end{pmatrix},
\end{equation}
where in addition to $n$ there are three free real parameters: two
parameters with the dimension of mass $m_a$ and $m_b$ which are proportional
to the reciprocal of the masses of the dominant and subdominant right-handed
neutrinos, and a relative phase $\eta$. A second version of this model has also
been proposed, based on an $S_4\times U(1)$ symmetry, where the second and
third rows and columns of the mass matrix are swapped~\cite{King:2016yvg}. In
this paper, we discuss both these versions for the case where $n=3$, with the
two versions of the model denoted as LSA and LSB;
\begin{align}
m^\nu_\text{LSA}&=m_a
\begin{pmatrix}
0 & 0 & 0 \\
0 & 1 & 1 \\
0 & 1 & 1
\end{pmatrix}
+m_be^{i\eta}
\begin{pmatrix}
1 & 3 & 1\\
3 & 9 & 3\\
1 & 3 & 1
\end{pmatrix},\label{eq:matrix_LSA}\\
m^\nu_\text{LSB}&=m_a
\begin{pmatrix}
0 & 0 & 0 \\
0 & 1 & 1 \\
0 & 1 & 1
\end{pmatrix}
+m_be^{i\eta}
\begin{pmatrix}
1 & 1 & 3\\
1 & 1 & 3\\
3 & 3 & 9
\end{pmatrix}.\label{eq:matrix_LSB}
\end{align}
Although, in the most minimal set-up, the relative phase $\eta$ is
a free parameter, it has been shown that in some models the presence of
additional $Z_3$ symmetries can fix the phase $e^{i\eta}$ to a cube root of
unity \cite{Ding:2013hpa},
%
%
with $\eta=2\pi/3$ the preferred value for LSA and $\eta=-2\pi/3$
for LSB as determined by current data \cite{Bjorkeroth:2014vha}.
This restriction gives the model greater predictivity by reducing the number of
free parameters to two, and we will give these cases special
attention while also showing some results for the case with $\eta$ left free.

Diagonalizing the mass matrices above leads to predictions for the
neutrino masses as well as the angles and phases of the unitary PMNS matrix,
$U_\text{PMNS}$, which describes the mixing between the three left-handed
neutrinos
\begin{equation} U_\text{PMNS}^Tm^\nu U_\text{PMNS}= \begin{pmatrix} m_1 & 0 &
0\\ 0 & m_2 & 0\\ 0 & 0 & m_3 \end{pmatrix},\\ \end{equation}
where $U_\text{PMNS}$ is defined by 
\begin{equation}
U_\text{PMNS}= \begin{pmatrix} \c12\c13 & \s12\c13 & \s13e^{-i\delta}\\
-\s12\c23-\c12\s13\s23e^{i\delta} & \c12\c23-\s12\s13\s23e^{i\delta} &
\c13\s23\\ \s12\s23-\c12\s13\c23e^{i\delta} & -\c12\s23-\s12\s13\c23e^{i\delta}
& \c13\c23 \end{pmatrix} \begin{pmatrix} e^{i\frac{\beta_1}{2}} & 0 & 0\\ 0 &
e^{i\frac{\beta_2}{2}} & 0\\ 0 & 0 & 1 \end{pmatrix} \end{equation}
with $\s{i}{j}=\sin\th{i}{j}$ and $\c{i}{j}=\cos\th{i}{j}$. All
of the parameters in this decomposition are therefore predicted in terms of the
2 (or 3) real parameters in \cref{eq:matrix_LSA,eq:matrix_LSB}.
Due to the minimal assumption of only two right-handed neutrinos,
the lightest neutrino is massless $m_1=0$ and the mass-squared differences,
which are the only combinations of masses accessible to neutrino oscillation
experiments, are predicted to be $\Dm21=m_2^2$ and
$\Dm31=m_3^2$. Of the remaining mixing parameters, \th12, \th13, \th23 and
$\delta$, are also experimentally accessible via neutrino
oscillation, while the Majorana phases $\beta_1$ and $\beta_2$
are not. 

As will be seen in more detail in the next section, due to their
similar forms, LSA and LSB make similar predictions. However, the process of
diagonalization reveals that the octant of \th23 is reversed, along with the
sign of $\delta$, while all other parameters are
unchanged. Changing the sign of $\eta$, however, also reverses the sign of
$\delta$ with no other effect, and so with the sign of $\eta$ not fixed by the
model the only physical difference between LSA and LSB is the octant of \th23.

\subsection{Sum rules of LS}

It has already been shown that, since the first column of the LS mixing matrix
$U_\text{PMNS}$ is equal to that of the tri-bimaximal mixing
matrix, LS (both LSA and LSB for all values of $\eta$) obeys the TM1 sum
rules~\cite{King:2015dvf,King:2016yvg}
\begin{align}
\tan\th12=\frac{1}{\sqrt{2}}\sqrt{1-3\s13^2},\qquad\sin\th12=&\frac{1}{\sqrt{3}}\frac{\sqrt{1-3\s13^2}}{\c13},\qquad\cos\th12=\sqrt{\frac{2}{3}}\frac{1}{\c13},\label{eq:sr1}\\
\cos{\delta}=&-\frac{\cot2\th23(1-5\s13^2)}{2\sqrt2\s13\sqrt{1-3\s13^2}},\label{eq:sr2}
\end{align}
where $\s{i}{j}=\sin\th{i}{j}$ and $\c{i}{j}=\cos\th{i}{j}$, and the forms in
\cref{eq:sr1} are equivalent.

For LSA with $\eta=\frac{2\pi}{3}$ or LSB with $\eta=-\frac{2\pi}{3}$, there
are several additional sum rules, which we discuss here for the first time. A
set of these additional sum rules can be derived using the fact that the only
two remaining input parameters $m_a$ and $m_b$ have dimensions of mass, so all
the mixing angles and phases must depend only on the ratio
$r\equiv\frac{m_b}{m_a}$. Exact expressions for the mixing angles and Dirac phase as a function of $r$ can be found in \cref{app:sr}, along with new exact sum rules derived using these expressions. These results make clear the difference between predictions of LSA and LSB; while \th13 and \th12 remain unchanged, $\cos2\th23$ and $\cos\delta$ differ by a change of sign.

An exact expression for the Jarlskog invariant $J$ was given
as~\cite{King:2015dvf,King:2016yvg}
\begin{equation} J= \s12\c12\s13\c13^2\s23\c23 \sin\delta=\mp
\frac{24m_a^3m_b^3(n-1)\sin\eta}{m_3^2m_2^2\Dm32}.  \label{eq:delta}
\end{equation}
with negative sign taken for LSA and positive for LSB.  For both LSA with
$\eta=\frac{2\pi}{3}$, and LSB with $\eta=-\frac{2\pi}{3}$ we find the new
relation
\begin{equation} m_2m_3=6m_am_b.  \end{equation}
Using this relation and inserting $n=3$ into \cref{eq:delta} leads to the new
relation for the Jarlskog invariant $J$ 
\begin{equation} J =-\frac{\sqrt{\Dm21\Dm31}}{3\sqrt{3}\Delta m^2_{32}}
\end{equation}
and hence the sum rule,
\begin{equation} \sin\delta=-\frac{\sqrt{\Dm21\Dm31}}{3\sqrt{3}\Delta
m^2_{32}\s12\c12\s13\c13^2\s23\c23}, \end{equation}
which is valid for both LSA with $\eta=\frac{2\pi}{3}$ and LSB with
$\eta=-\frac{2\pi}{3}$.

\section{Probing LS with existing data}\label{sec:data}

Existing measurements of the neutrino mixing parameters have been shown to be
in good agreement for CSD($n$) for the $n=3$ case \cite{Bjorkeroth:2014vha}.
The best-fit value of $\eta$ is found to be close to $\pm\frac{2\pi}{3}$, with
the positive sign for LSA and the negative sign for LSB, which has been
theoretically motivated as one of the cube roots of unity required due to an
additional $Z_3$ symmetry as part of a larger GUT model \cite{King:2015dvf}.
In this section, we study both the case where $\eta$ is fixed by symmetry and
the case where it is left as a free parameter of the theory.

\subsection{Predictions of oscillation parameters with fixed \texorpdfstring{$\eta=\pm 2\pi/3$}{eta=2pi/3}}\label{sec:pred}

\begin{figure}[ht]
	\centering\includegraphics[width=\textwidth]{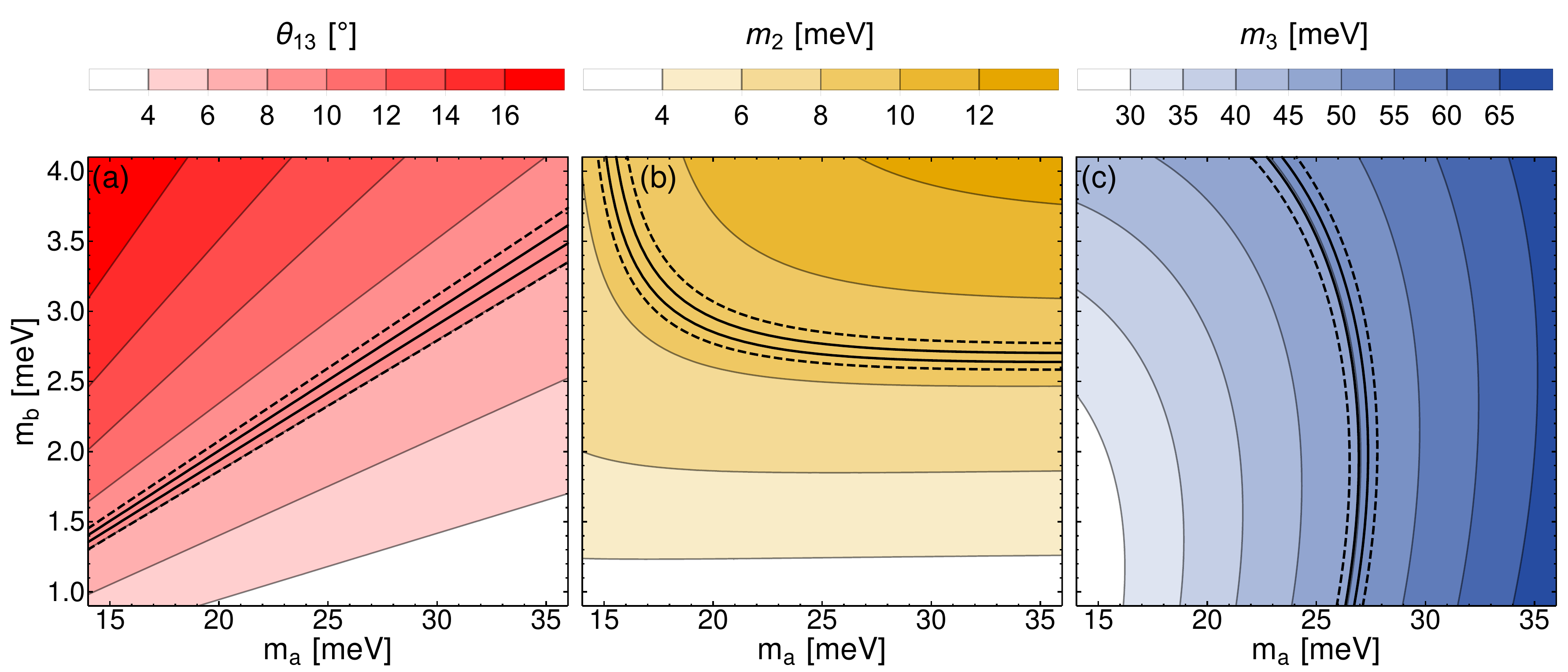}
	\caption{Predicted values from LSA with $\eta=\frac{2\pi}{3}$ (or LSB with $\eta=-\frac{2\pi}{3}$)
of oscillation parameters depending on the input parameters $m_a$ and $m_b$. Regions corresponding to the experimentally determined $1\sigma$ (solid lines) and $3\sigma$ (dashed lines) ranges for each parameter are also shown.}
	\label{fig:params-ma-mb}
\end{figure}

In the $n=3$ case of LSA with $\eta=\frac{2\pi}{3}$ (or LSB with
$\eta=-\frac{2\pi}{3}$), all neutrino masses, mixing angles and phases are
fully determined from the two remaining parameters $m_a$ and $m_b$ and the
three most precisely measured of these parameters, $\th13$, $\Dm31$ and
$\Dm21$, currently provide the strongest test of the LS model.
\Cref{fig:params-ma-mb} shows how these parameters vary in the $m_a-m_b$ plane,
along with the regions corresponding to the $1\sigma$ and $3\sigma$ ranges for
these parameters from the NuFit 3.0 (2016) global fit
\cite{Esteban:2016qun}, assuming normal mass ordering and a lightest
neutrino mass of $m_1=0$. The SD proposal requires $m_a$ to be significantly
larger than $m_b$ and for this portion of the parameter space the approximate
proportionality relations of $m_2\sim m_b$ and $m_3\sim m_a$ can be seen,
verifying the approximations previously derived in \cite{King:2015dvf}.

\begin{figure}[ht]
	\centering\includegraphics[width=0.6\textwidth]{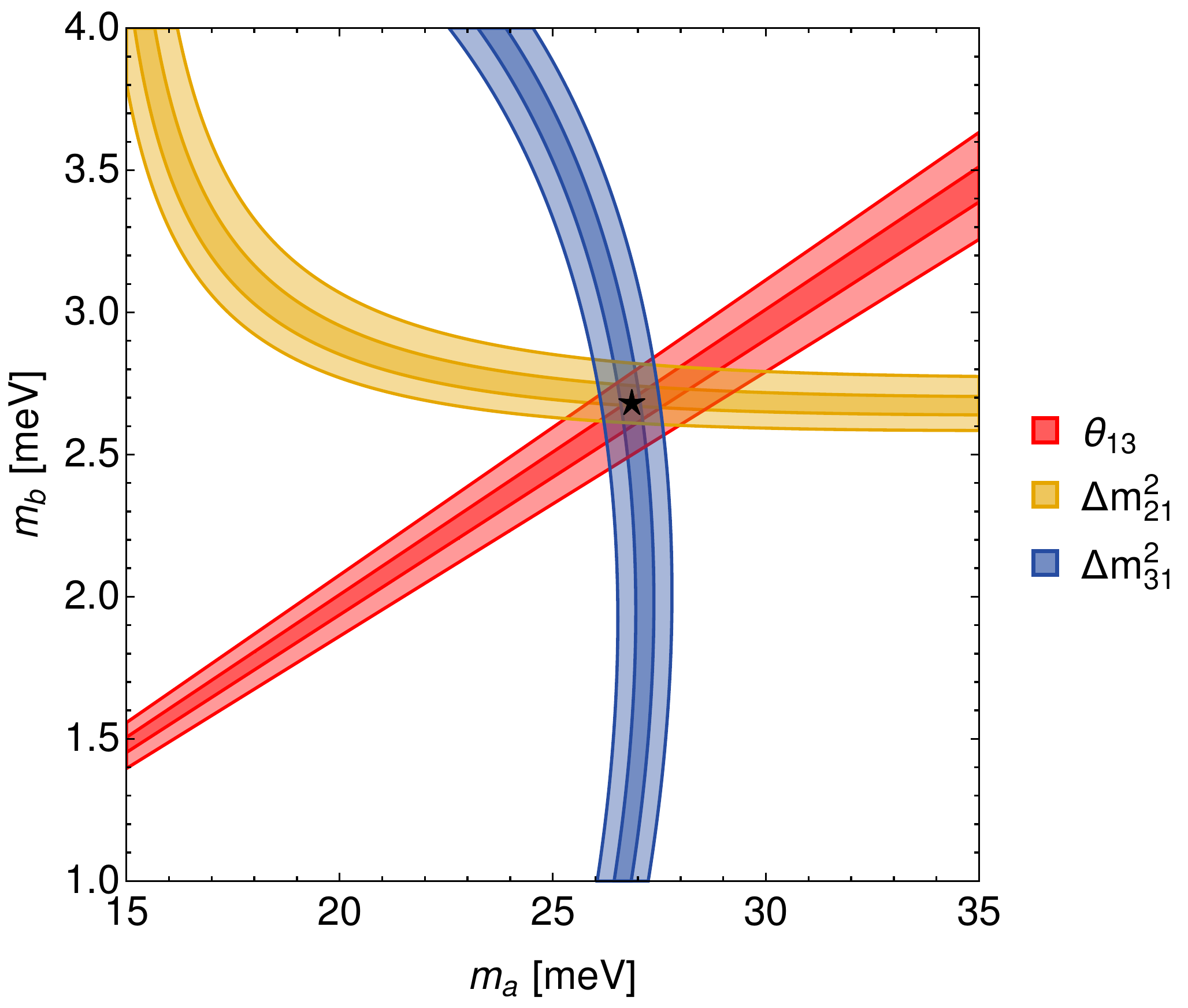}
	\caption{Regions in the $m_a$-$m_b$ plane with fixed $\eta=2\pi/3$ ($\eta=-2\pi/3$) for LSA (LSB) corresponding to the experimentally determined $1\sigma$ and $3\sigma$ ranges for $\th13$, $\Dm21$ and $\Dm31$.}
	\label{fig:ma-mb-contours}
\end{figure}

\begin{figure}[ht]
	\centering\includegraphics[width=\textwidth]{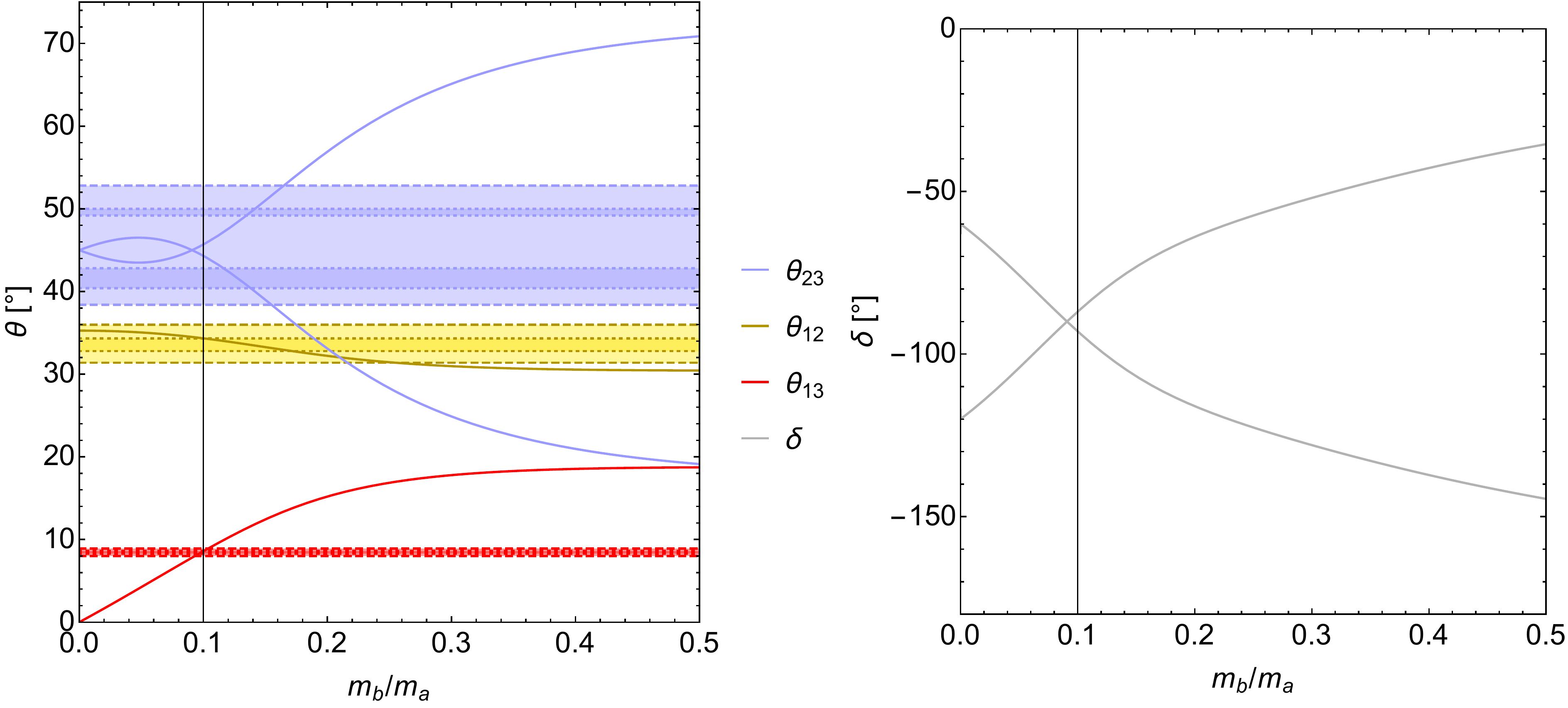}
	\caption{Predicted values from LS with fixed $\eta=2\pi/3$ ($\eta=-2\pi/3$) for LSA (LSB) of the mixing angles and delta as a function of the ratio $m_b/m_a$. Horizontal bands show the experimentally determined $1\sigma$ and $3\sigma$ ranges for each parameter. A reference point giving a good prediction for all parameters is shown at $r=m_b/m_a=0.1$.}
	\label{fig:params-ratio}
\end{figure}

Even at $1\sigma$ the three allowed regions coincide at a single point, as can be seen in \cref{fig:ma-mb-contours}, and so this benchmark point can be used to make predictions of the remaining angles $\th12$ and $\th23$ and the Dirac phase $\delta$.
As described in \cref{sec:csd} these parameters, along with $\th13$, depend only on the ratio $r=m_b/m_a$; this dependence, given by the relations in \cref{eq:params-ratio}, is shown in \cref{fig:params-ratio}, with the $1\sigma$ and $3\sigma$ NuFIT 3.0 ranges and reference point at $m_b/m_a=0.1$. For \th23 and $\delta$, the predictions of both LSA and LSB are shown. At this point it can be seen that while both $\th13$ and $\th12$ lie within their $1\sigma$ ranges, $\th23$ lies just outside its $1\sigma$ range, and a prediction on the value of the Dirac phase is made of $\delta\simeq-90^\circ$.

\begin{figure}[ht]
	\centering\includegraphics[width=\textwidth]{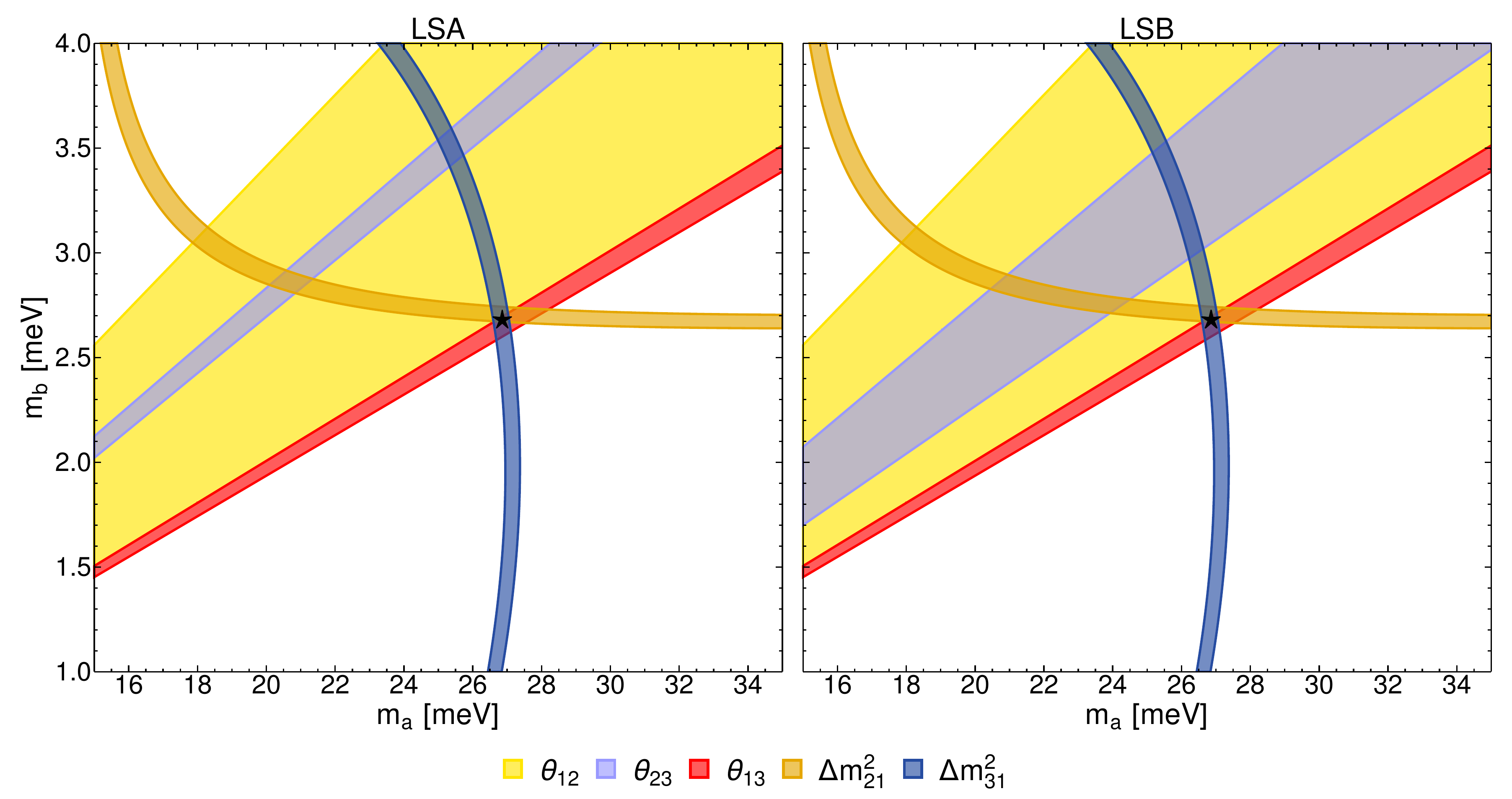}
	\caption{Regions in the $m_a$-$m_b$ plane with fixed $\eta=2\pi/3$ ($\eta=-2\pi/3$) for LSA (LSB) corresponding to the experimentally determined $1\sigma$ ranges for solar and reactor mixing angles and mass-squared differences. The \th23 regions shown are in tension with other measurements, however, extending to $2\sigma$ these regions become far larger, covering the entire parameter space shown in these plots.}
	\label{fig:ma-mb-contours2}
\end{figure}

Combining these results for all parameters which have been experimentally
measured, displayed together in \cref{fig:ma-mb-contours2}, it is seen that
the prediction for \th12 lies just within current bounds.
However, there is tension at the $1\sigma$ level for \th23, due to
the allowed regions of LS parameter space requiring values close to maximal,
while current data points towards larger deviations from the maximal value. The experimental measurements of \th23 do not yet give consistent indications of its value; while the latest results from NO$\nu$A disfavour maximal mixing at $2.5\sigma$~\cite{nova:nu2016}, results from T2K remain fully compatible with maximal \th23~\cite{t2k:ichep2016}. As a result, while the combined fit for \th23 is in tension with the LS models at $1\sigma$, the allowed range at $2\sigma$ is far wider, crossing both octants and the maximal value of $45^\circ$, including the values preferred by the LS model\footnote{For a more detailed discussion of the current status of experimental measurements of \th23, see \cite{Esteban:2016qun}}.

\subsection{Predictions of oscillation parameters with \texorpdfstring{$\eta$}{eta} as a free parameter}

\begin{figure}[ht]
	\centering\includegraphics[width=\textwidth]{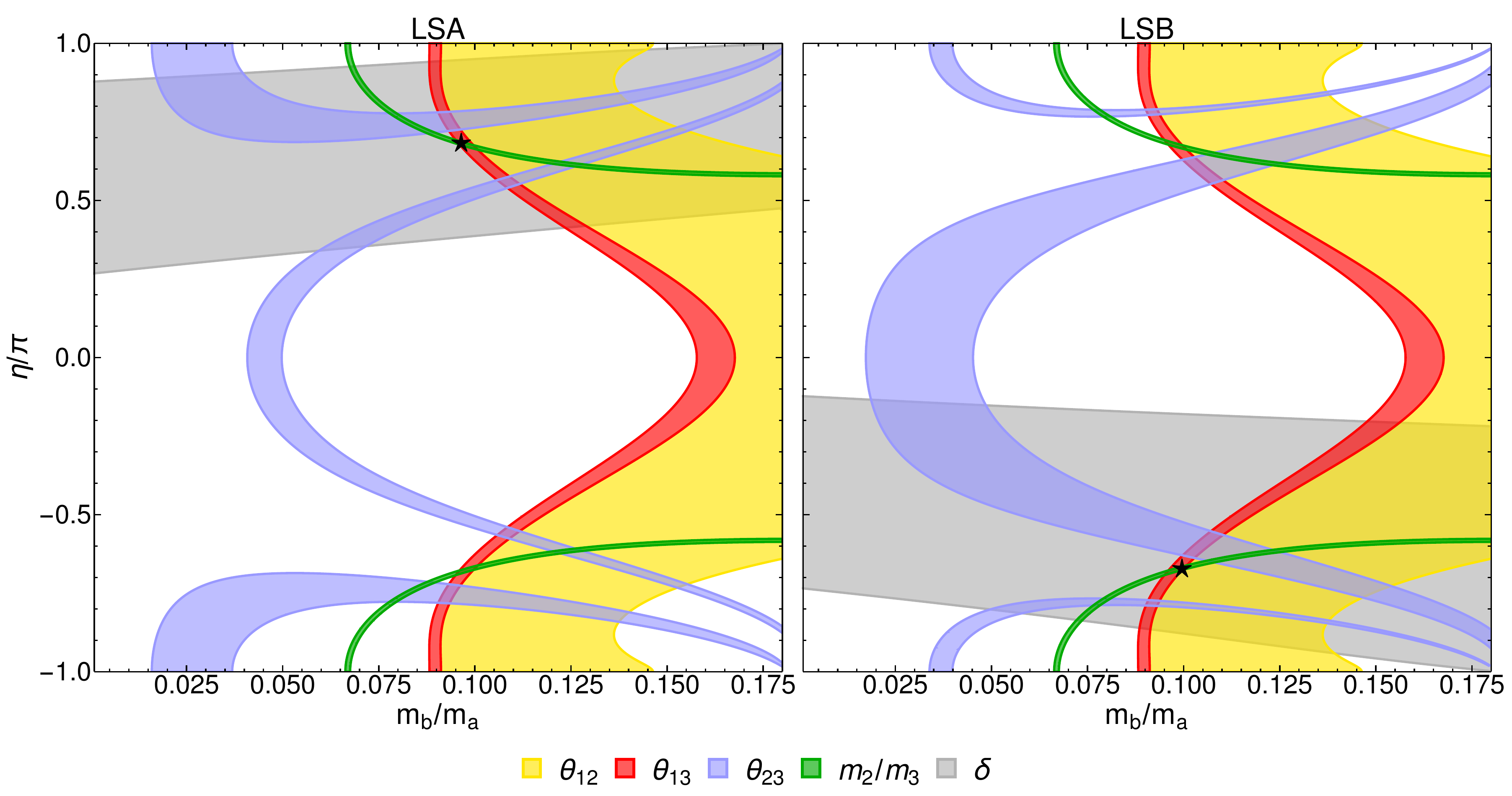}
	\caption{Regions in the $m_b/m_a$-$\eta$ plane corresponding to the experimentally determined $1\sigma$ ranges for all mixing angles, $\delta$ and the ratio of neutrino masses $m_2/m_3$ for LSA (left panel) and LSB (right panel).}
	\label{fig:r-eta-contours}
\end{figure}

In the versions of the LS models with $\eta$ as an additional free parameter,
the mixing angles and phases now depend on both the ratio $r=m_b/m_a$ and
$\eta$. The masses $m_3$ and $m_2$ depend on all three input
parameters; however, their ratio $m_2/m_3$ (and
therefore the ratio $\Dm21/\Dm31$) will depend only on $r$ and $\eta$. As
previously, the strongest contraints come from the very precise measurements of
\th13 and the mass-squared differences \Dm21 and \Dm31.
\Cref{fig:r-eta-contours} shows the regions corresponding to the $1\sigma$
ranges for all the mixing angles, $\delta$ and $m_2/m_3$, where we see that all
the five regions come close to overlapping around $\eta=\pm 2\pi/3$ for LSA and
LSB, respectively. That two input parameters should give a good description of
five observables, within their one sigma errors, is ostensibly a remarkable
achievement, indeed perhaps better than might be expected on statistical
grounds. However, due to the very tight constraints on $\eta$
from \th13 and $m_2/m_3$, we still find some tension with the
value of \th23 even when allowing $\eta$ to vary. As with the case with $\eta$ fixed, this tension exists only at the $1\sigma$ level, where close to maximal \th23 is excluded.

\subsection{Fitting LS models to global fit data }\label{sec:fit}

In order to provide a more concrete measure of the agreement between the
predictions of the model and existing data, as well as to make
further predictions of the less well measured parameters, we have performed a
$\chi^2$ fit to the four cases discussed above: LSA and LSB with $\eta$ fixed
and free. As a proxy for the full data sets of previous
experiments, our fits use the results of the NuFIT 3.0 global analysis
\cite{Esteban:2016qun}. This analysis combines the latest results (as of fall 2016) of solar, atmospheric, long baseline accelerator, and long, medium and short baseline reactor neutrino experiments, to obtain a combined fit to the six standard neutrino oscillation parameters.
We use the $\chi^2$ data provided by NuFIT, for the case where normal mass ordering is assumed, combining both the 1D $\chi^2$ data for each mixing parameter with the 2D $\chi^2$ data to include correlations between parameter measurements
\begin{equation}
\chi^2_\text{Fit}(\Theta)=\sum_{\theta_i\in\Theta}\chi^2_\text{1D}(\theta_i)+\sum_{\theta_i\neq\theta_j\in\Theta}\left(\chi^2_\text{2D}(\theta_i,\theta_j)-\chi^2_\text{1D}(\theta_i)-\chi^2_\text{1D}(\theta_j)\right), \label{eq:fit}
\end{equation}
where the first sum in this expression combines each of the 1D $\chi^2$ data into a
first approximation of the full 6D $\chi^2$ while the second sum provides
corrections to this coming from the 2D correlations between each pair of
parameters. 

We then apply this result first to the standard mixing case, then to the LS model case as follows:

\begin{itemize}
\item For the case of standard mixing $\Theta=\Theta_\text{PMNS}\equiv\left\{\th12,\th13,\th23,\Dm21,\Dm31,\delta\right\}$
and we simply combine the NuFIT 3.0 results as shown above, in order to include correlations,
and use it to calculate $\chi^2\left(\Theta_\text{PMNS}\right)\equiv \chi^2_\text{Fit}(\Theta)$ for this case.
\item For the LS model 
we use instead $\Theta=\Theta_{LS}\equiv\left\{m_a,m_b,\eta\right\}$ (or $\Theta_\text{LS}=\left\{m_a,m_b\right\}$ when fitting with $\eta$ fixed),
which is then minimised over the LS parameter space using the
analytic relations to calculate standard mixing parameters from LS parameters,
and hence calculate $\chi^2\left(\Theta_\text{LS}\right) \equiv \chi^2_\text{Fit}(\Theta)$ for this case.
\end{itemize}

Our test statistic for a particular LS model is then given by:
\begin{equation}
\sqrt{\Delta\chi^2}=\sqrt{\min_{\Theta_\text{LS}}\left[\chi^2\left(\Theta_\text{LS}\right)\right]-\min_{\Theta_\text{PMNS}}
\left[\chi^2\left(\Theta_\text{PMNS}\right)\right]}.
\end{equation}
We have verified through Monte-Carlo calculations that Wilk's theorem holds for
this statistic, \ie it is approximately distributed according to a chi-squared
distribution.

The best fit LSA and LSB points for fits with $\eta$ left free or
with $\eta$ fixed at $\frac{2\pi}{3}$ are given in \cref{tab:bfp}. 
The number of degrees of freedom (d.o.f.) is either 3 or 4, which is just the difference between the
number of observables (which we take to be the parameters in $\Theta_\text{PMNS}$)
and the number of LS parameters (namely the parameters in $\Theta_\text{LS}$,
which is either 3 or 2,
depending on whether $\eta$ is free or fixed). 
For LSA we
find a best fit with $\Delta\chi^2=4.1$ (3 degrees of freedom) with $\eta$ free
and $\Delta\chi^2=5.6$ (4 degrees of freedom) fixing $\eta=\frac{2\pi}{3}$,
while for LSB we find better fits, with $\Delta\chi^2=3.9$ (3 degrees of
freedom) and $\Delta\chi^2=4.5$ (4 degrees of freedom) for $\eta$ free and
$\eta=-\frac{2\pi}{3}$ respectively. 

\begin{figure}[t]
\centering\includegraphics[width=\textwidth]{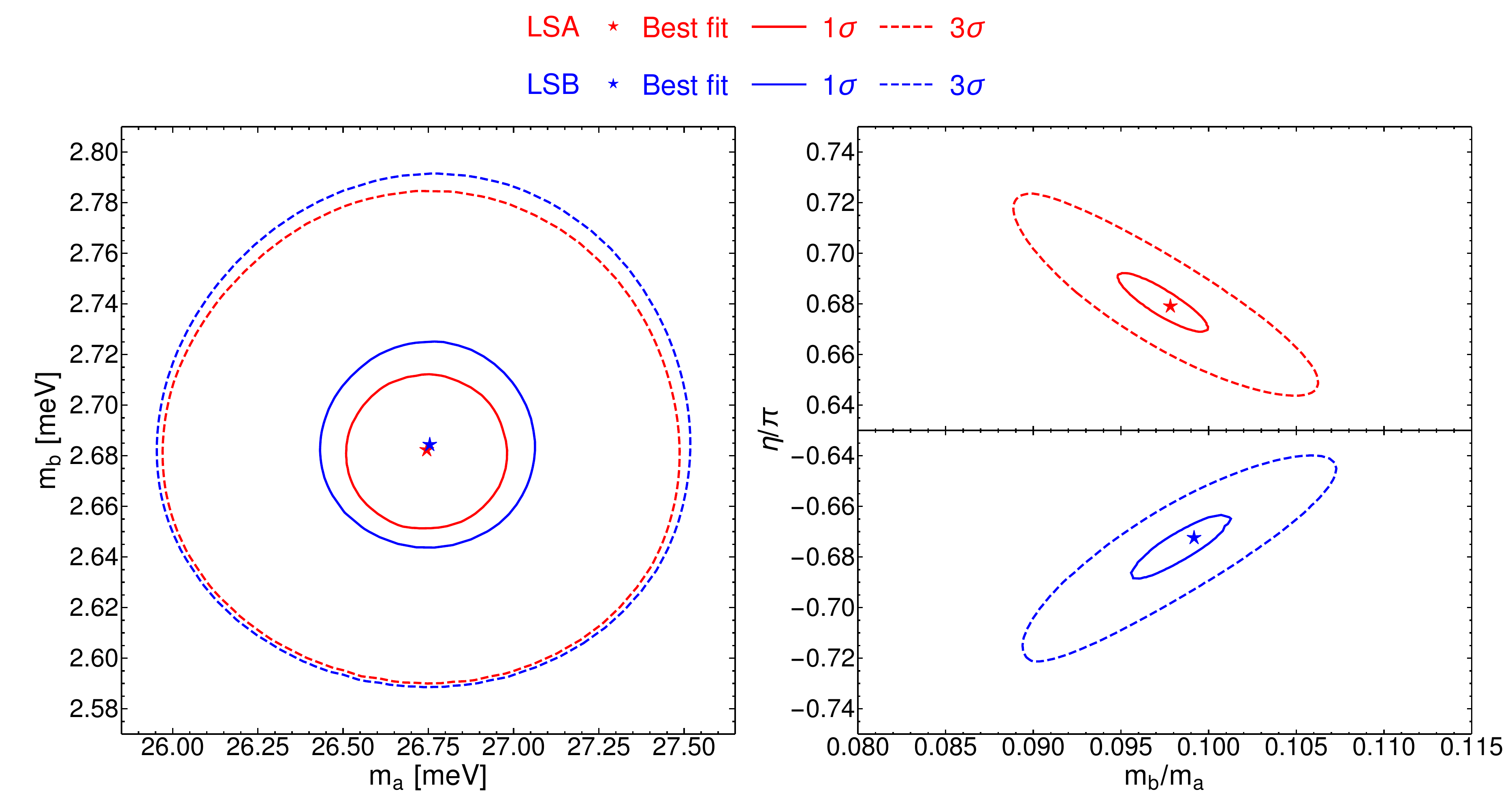}
\caption{Results of the fits to LS of the NuFIT 3.0 (2016) global neutrino oscillation data. Left: LS fit with fixed $\eta=2\pi/3$ ($\eta=-2\pi/3$) for LSA (LSB). Right: LS fit with $\eta$ as a free parameter.}
\label{fig:fit}
\end{figure}

\Cref{fig:fit} shows the best fit points
with $1\sigma$ and $3\sigma$ contours of the fits in the $m_a-m_b$ plane for
fixed $\eta$ and in the $r-\eta$ plane for free $\eta$. 
The significance at which a LS model is allowed is determined from the distribution of the $\Delta\chi^2$ test statistic,
where $N\sigma$
has been calculated assuming the that Wilks' theorem applies.
Note that despite LSA
predicting values of \th23 which lie outside its individual $1\sigma$ range
reported by NuFIT 3.0, there are still regions not excluded at $1\sigma$. This
is due to the high predictivity of the model; by predicting many parameters
from few input parameters there is a greater chance that one of these may lie
outside its experimentally determined range. Statistically, this comes from the
increased number of degrees of freedom of the $\chi^2$-distribution which
approximates our test statistic $\Delta\chi^2$.

\begin{table}[t]
\centering\begin{tabular}{lrrrrr}
		\toprule
		 & \multicolumn{2}{c}{LSA} & \multicolumn{2}{c}{LSB} & \multicolumn{1}{c}{NuFIT 3.0} \\
		 & \multicolumn{1}{c}{$\eta$ free} & \multicolumn{1}{c}{$\eta$ fixed} & \multicolumn{1}{c}{$\eta$ free} & \multicolumn{1}{c}{$\eta$ fixed} & \multicolumn{1}{c}{global fit} \\
		\midrule
		$m_a$ [meV] & 27.19 & 26.74 & 26.95 & 26.75 & \\
		$m_b$ [meV] & 2.654 & 2.682 & 2.668 & 2.684 & \multicolumn{1}{c}{---} \\
		$\eta$ [rad] & $0.680\pi$ & $2\pi/3$ & $-0.673\pi$ & $-2\pi/3$ &  \\
		\midrule
		$\th12$ [$^\circ$] & 34.36 & 34.33 & 34.35 & 34.33 & $33.72^{+0.79\phantom{0}}_{-0.76}$ \\
		$\th13$ [$^\circ$] & 8.46 & 8.60 & 8.54 & 8.60 & $8.46^{+0.14\phantom{0}}_{-0.15}$ \\
		$\th23$ [$^\circ$] & 45.03 & 45.71 & 44.64 & 44.28 & $41.5^{+1.3\phantom{00}}_{-1.1}$ \\
		$\delta$ [$^\circ$] & -89.9 & -86.9 & -91.6 & -93.1 & $-71^{+38\phantom{.00}}_{-51}$ \\
		$\Dm21$ [$10^{-5}\text{eV}^2$] & 7.499 & 7.379 & 7.447 & 7.390 & $7.49^{+0.19\phantom{0}}_{-0.17}$ \\
		$\Dm31$ [$10^{-3}\text{eV}^2$] & 2.500 & 2.510 & 2.500 & 2.512 & $2.526^{+0.039}_{-0.037}$ \\
		\midrule
		$\Delta\chi^2$\,/\,d.o.f & 4.1\,/\,3 & 5.6\,/\,4 & 3.9\,/\,3 & 4.5\,/\,4 & \multicolumn{1}{c}{---} \\
\bottomrule
\end{tabular}
\caption{Results of our fit of existing data to LSA and LSB with $\eta$ left free and for $\eta=\frac{2\pi}{3}$ for LSA and $\eta=-\frac{2\pi}{3}$ for LSB. The results of the NuFIT 3.0 (2016) global fit to standard neutrino mixing are shown for the normal ordering case for comparison.}
\label{tab:bfp}
\end{table}

\begin{figure}[t]
	\centering
	\sublabel{fig:fitpanel1.a}
	\sublabel{fig:fitpanel1.b}
	\sublabel{fig:fitpanel1.c}
	\sublabel{fig:fitpanel1.d}
	\sublabel{fig:fitpanel1.e}
	\sublabel{fig:fitpanel1.f}
	\includegraphics[width=\textwidth]{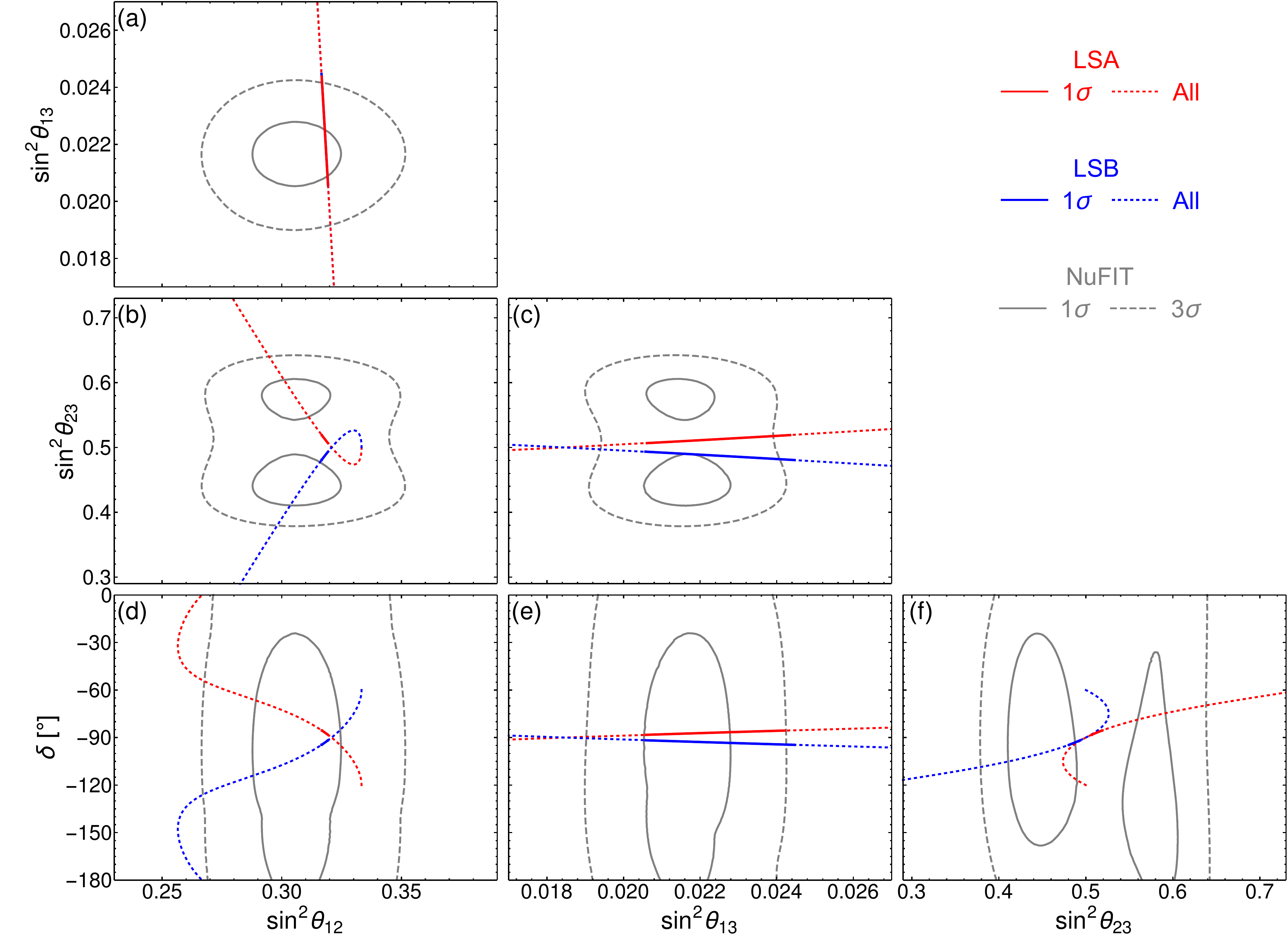}
	\caption{Allowed values for LSA (red) and LSB (blue) with $\eta=2\pi/3$ and $\eta=-2\pi/3$ respectively, showing all possible values (dotted) and the $1\sigma$ range (solid). These lines of allowed solutions correspond to the sum rules in \cref{eq:sr1,eq:sr}, or combinations thereof. Also shown are the $1\sigma$ (solid) and $3\sigma$ (dashed) regions from the NuFIT 3.0 2016 global fit (grey).}
	\label{fig:fitpanel1}
\end{figure}

Our fit can also be used to identify the regions of standard neutrino mixing
parameter space predicted by LS, once existing data has been taken into
account. This corresponds to mapping the regions of LS input parameter space
allowed by our fit onto the standard mixing parameter space. 
\Cref{fig:fitpanel1} shows the predictions of LS (for the fixed
$\eta$ case) in the planes made from each pair of mixing angles
and $\delta$. Since these values all depend only on the single parameter $r$,
the predictions of LS form lines of allowed solutions in each
plane, corresponding to sum-rules between the oscillation parameters.  For
example, \cref{fig:fitpanel1.a} corresponds to the TM1 sum rule in
\cref{eq:sr1}, while \cref{fig:fitpanel1.b,fig:fitpanel1.c,fig:fitpanel1.d,fig:fitpanel1.e,fig:fitpanel1.f} correspond to those in \cref{eq:sr} or to
combinations of these sum rules. It can be seen that very strong restrictions
are placed on the allowed values of the less well measured parameters, \th12,
\th23 and $\delta$. For the remaining angle, \th13, around two thirds of the
NuFIT 3.0 range remains viable in LS. 

\begin{figure}[t]
	\centering
	\sublabel{fig:fitpanel2.a}
	\sublabel{fig:fitpanel2.b}
	\sublabel{fig:fitpanel2.c}
	\sublabel{fig:fitpanel2.d}
	\sublabel{fig:fitpanel2.e}
	\sublabel{fig:fitpanel2.f}
	\sublabel{fig:fitpanel2.g}
	\sublabel{fig:fitpanel2.h}
	\sublabel{fig:fitpanel2.i}
	\includegraphics[width=\textwidth]{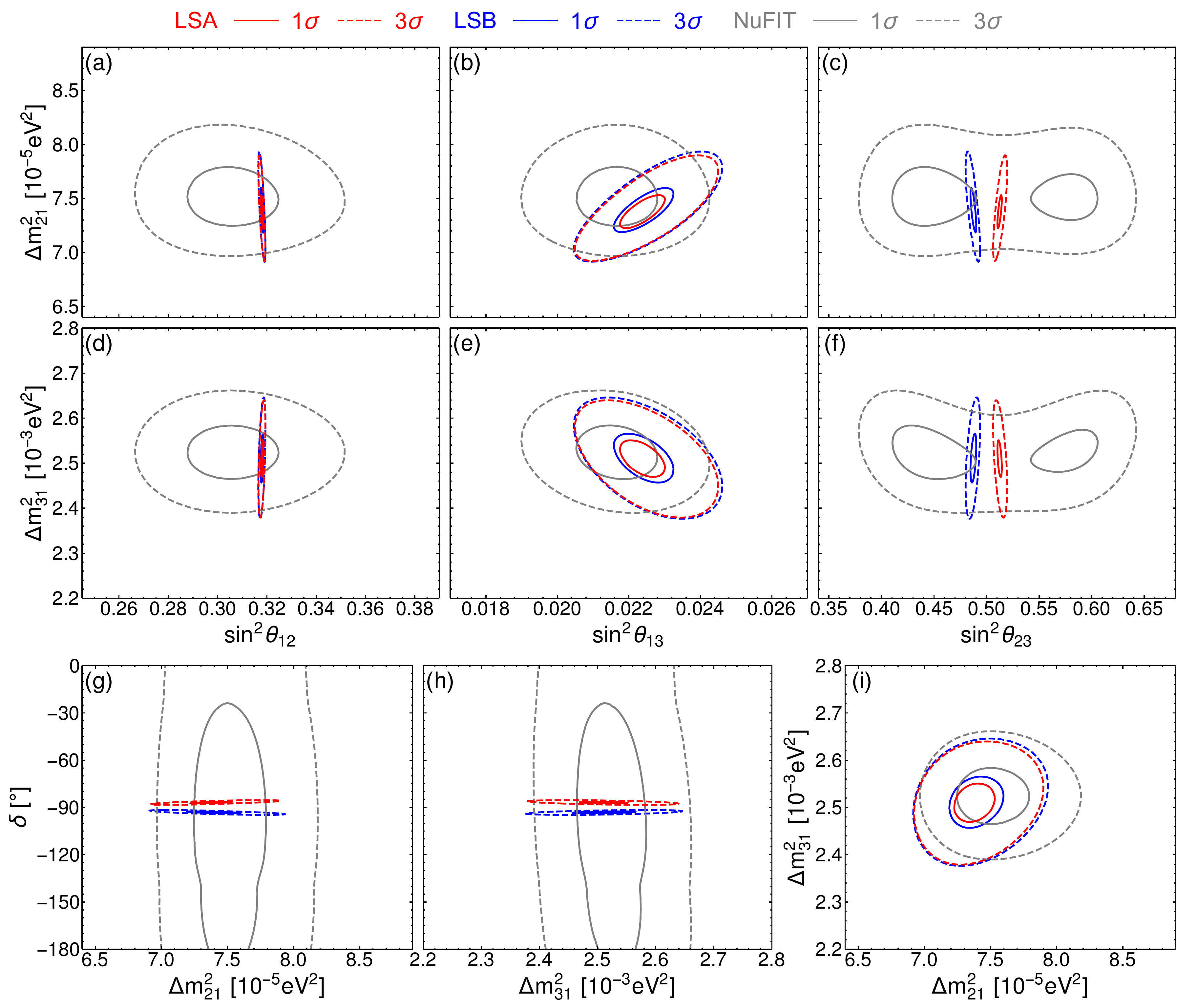}
	\caption{Allowed $1\sigma$ (solid) and $3\sigma$ (dashed) regions for LSA (red) and LSB (blue) with $\eta=2\pi/3$ and $\eta=-2\pi/3$ respectively. Also shown are the current allowed regions from the NuFIT 3.0 2016 global fit (grey).}\label{fig:fitpanel2}
\end{figure}

\Cref{fig:fitpanel2} shows the allowed regions of parameter space for pairs of
variables including the mass-squared differences. In these plots,
as the mass-squared differences can depend on both $m_a$ and $m_b$
independently, we see regions of allowed values instead of lines. For each of
these planes, any point will fully determine both input parameters $m_a$ and
$m_b$, and so these contours correspond exactly to the equivalent regions shown
in \cref{fig:fit}. In addition to the tight constraints on \th12, \th23 and
$\delta$ already mentioned, in \cref{fig:fitpanel2.b,fig:fitpanel2.e} it can
be seen that the allowed range of \th13 is correlated with that of both \Dm21
and \Dm31, suggesting that combining future measurements of these parameters
could provide a better probe of LS than the individual parameter
measurements alone.  The ability of future experiment to exclude the model then
depends on both the predictions of the model seen here, combined with the
sensitivity of experiments to measurements of the parameters in the region of
interest predicted by LS, which is the focus of the next section.


\section{Sensitivity of future experiments}\label{sec:sens}

In order to understand the potential for future experiments to exclude the LS
models, we have performed simulations of a combination of accelerator and
reactor experiments, modelling the experimental data expected over
the next two decades. We have used the General Long Baseline Experiment
Simulator (GLoBES) libraries~\cite{Huber:2004ka,Huber:2007ji} to simulate
future experiments and to fit the simulated data to both standard mixing and
the LS models. In all our simulations we assume that the mass ordering is known
to be normal ordering, as this is a requirement of the LS models; a measurement
of inverted ordering would immediately exclude the models.

\subsection{Future neutrino oscillation experiments}\label{sec:sim}

Our combination of experiments include detailed simulations of the
T2HK and DUNE long-baseline accelerator experiments, which aim to
provide precision measurements of \Dm31, \th23 and $\delta$, together with
basic constraints on \th13 from the Daya Bay short baseline reactor experiment
and on \th12 and \Dm21 from the JUNO and RENO-50 medium baseline
reactor experiments. We will now briefly recap the salient
features of these experiments and our treatment of them.

\subsubsection*{T2HK}

The Tokai to Hyper-Kamiokande (T2HK) experiment is a proposed
long-baseline accelerator neutrino experiment using the Hyper-Kamiokande detector,
a megatonne scale water Cherenkov detector to be constructed near to the
Super-Kamiokande detector in Kamioka, Japan~\cite{HKDR}. The standard design is for two tanks
to be built, each with 258~kt (187~kt) of total (fiducial) volume. The tanks are to
be built in a staged process with the second tank constructed and commissioned after
the first, such that the second begins to take data six years after the first. The
water Cherenkov technique is capable of detecting the (anti-)muons and electrons
(positrons) produced in (anti-)neutrino interactions, with the ability to
distinguish the charged leptons' flavours but not their charge. The detector
would be used to observe neutrinos from (amongst other sources) an upgraded
version of the T2K neutrino beam produced at J-PARC in Tokai, $L=295$~km from the
detector. The 1.3~MW beam, produced from a 30~GeV protons, is directed
2.5$^\circ$ away from the detector in order to provide a narrow energy spectrum
at the far detector peaked around the first atmospheric neutrino oscillation maximum 
for $\Dm31\sim 2.5\times 10^{-3}$ eV$^2$ and $E=0.6$~GeV.
Either $\nu_\mu$ or $\bar\nu_\mu$ can be produced as the principle component of
the beam, such that the oscillation probabilities $P(\nu_\mu\to\nu_e)$,
$P(\bar\nu_\mu\to\bar\nu_e)$, $P(\nu_\mu\to\nu_\mu)$ and
$P(\bar\nu_\mu\to\bar\nu_\mu)$ can all be measured.  While the main goal of
T2HK is to search for CP symmetry violation by observing a non-CP
conserving value of $\delta$, precision measurements of \th23 and the magnitude
of \Dm31 will also be made~\cite{Abe:2015zbg}.

\subsubsection*{DUNE}

The Deep Underground Neutrino Experiment~\cite{Acciarri:2016crz} (DUNE) is a
proposed long-baseline accelerator experiment, which differs from the
T2HK experiment through its longer baseline and higher energy
wide-band beam. The experiment will use a new neutrino beam sourced at
Fermilab, directed towards a large liquid argon detector in Sanford, $L=$1300~km
from the beam source. The 40~kt LArTPC detector is able to detect both the charged
leptons and the hadrons produced from muon and electron (anti-)neutrino interactions,
with strong particle identification and energy reconstruction capabilities. The standard
design is for a 1.07~MW $\nu_\mu$ or $\bar\nu_\mu$ beam produced from 80~GeV protons,
with an on-axis design to produce a wide energy spectrum spanning $E=$0.5 to 5~GeV,
allowing observations of the $\nu_e$ appearance spectrum around 
the first atmospheric neutrino oscillation maximum 
for $\Dm31\sim 2.5\times 10^{-3}$ eV$^2$. While measuring the same oscillation channels as
T2HK, the wider band beam with longer baseline provides complementary
information on the value of $\delta$ as well as measurements of \th23 and, due
to the matter effects from the longer baseline, both the sign and the magnitude
of \Dm31~\cite{Acciarri:2015uup}.

\subsubsection*{Short baseline reactor experiments}

By observing the oscillations of the $\bar\nu_e$ produced in nuclear reactors,
short baseline reactor neutrino experiments are able to measure
the mixing angle \th13 with particularly high accuracy. The Daya Bay
experiment~\cite{Guo:2007ug} currently has the most precise measurement of this
parameter with the aim to achieve a precision on $\sin^2\th13$ of better than
3\%~\cite{Cao:2016vwh}. The experiment measures anti-neutrinos produced in six
nuclear reactors in south China. A total of eight 20~t liquid scintillator
detectors are used; two are located at each of two near detector sites and four
at a far detector site $L=$1.5 to 1.9~km from the reactors near 
the first atmospheric neutrino oscillation maximum 
for $\Dm31\sim 2.5\times 10^{-3}$ eV$^2$, given the low nuclear energy of the neutrino
beam $E\sim $ few MeV.
Results of
the Double Chooz \cite{Ardellier:2004ui} and RENO
\cite{Ahn:2010vy, Kim:2014rfa} short baseline reactor experiments
also contribute to the precision obtained on \th13 combined with the Daya Bay
result. Although DUNE and T2HK will also measure this parameter
with high precision, the measurement of the short
baseline reactor programme by that time is expected to be at least as
precise, and will provide a measurement independent of the other
parameters which influence the appearance channel at
long-baseline accelerator experiments.

\subsubsection*{Medium baseline reactor experiments}

The Jiangmen Underground Neutrino Observatory~\cite{Djurcic:2015vqa} (JUNO)
and the future plans of the Reactor Experiment for Neutrino
Oscillation (RENO-50)~\cite{Kim:2014rfa} are medium baseline reactor neutrino
experiments which, like the Daya Bay experiment, will observe the oscillations
of electron anti-neutrinos produced in nuclear reactors. The JUNO experiment
will use a 20~kt liquid scintillator detector approximately $L=$53~km from two
planned nuclear reactors in southern China, while RENO-50 will use
an 18~kt liquid scintillator detector approximately $L=$50~km from a nuclear
reactor in South Korea. Given the low nuclear energy of the neutrino
beam $E\sim $ few MeV, these longer baselines correspond to 
the first solar neutrino oscillation maximum 
for $\Dm21\sim 7.5\times 10^{-5}$ eV$^2$, where the higher frequency atmospheric
oscillations appear as wiggles.
Thus the longer baseline than at Daya Bay gives greatest
sensitivity to a different set of oscillation parameters, in particular \th12
and \Dm21. The precision on the measurements of both $\sin^2\th12$ and \Dm21 is
expected to reach 0.5\%~\cite{Djurcic:2015vqa,Kim:2014rfa}.

\subsubsection*{Details of experimental simulation}

\begin{table}[t]
\centering\begin{tabular}{lll}
\toprule
Experiment & Parameter & Precision\\
\midrule
Short baseline reactor & $\sin^2\th13$ & 3\%\\
Medium baseline reactor & $\sin^2\th12$ & 0.5\%\\
Medium baseline reactor & \Dm21 & 0.5\%\\
\bottomrule
\end{tabular}
\caption{Precision of oscillation parameter measurements made by reactor experiments which we have used as constraints in our simulations.}
\label{tab:precision}
\end{table}

We have used complete simulations of the latest designs for both DUNE and
T2HK where we have assumed both experiments run for 10 years. Full
details of the GLoBES implementations we have used can be found
in~\cite{synergy}. For the short and medium baseline reactor experiments, we have included
basic constraints on the values of $\sin^2\th13$, $\sin^2\th12$ and \Dm21.
Since these measurements are expected to be approximately independent of other
parameters we have implemented these constraints as simple Gaussian
measurements with a mean of the true simulated value and error as given in
\cref{tab:precision}.

\subsection{Statistical method}\label{sec:stat}

To determine the statistical significance with which the LS model could be excluded based on simulated data, we perform a minimum-$\chi^2$ fit to both standard three neutrino mixing and to the LS model. 
As in section~\ref{sec:fit}, 
for the case of standard mixing we use $\Theta=\Theta_\text{PMNS}\equiv\left\{\th12,\th13,\th23,\Dm21,\Dm31,\delta\right\}$, while for LS we use $\Theta=\Theta_{LS}\equiv\left\{m_a,m_b,\eta\right\}$ (or $\Theta_\text{LS}=\left\{m_a,m_b\right\}$ when fitting with $\eta$ fixed). Our test statistic for the significance to exclude the LS model is then given by
\begin{equation}
\sqrt{\Delta\chi^2}=\sqrt{\min_{\Theta_\text{LS}}\left[\chi^2\left(\Theta_\text{LS}\right)\right]-\min_{\Theta_\text{PMNS}}\left[\chi^2\left(\Theta_\text{PMNS}\right)\right]}.
\end{equation}
The significance at which LS is excluded is then determined from the distribution of the $\Delta\chi^2$ test statistic; where we give sensitivities in terms of $N\sigma$, this quantity has been calculated assuming the that Wilks' theorem applies. Wilks' theorem states that when comparing nested models, the $\Delta\chi^2$ test statistic is a random variable asymptotically distributed according to the $\chi^2$-distribution with the number of degrees of freedom equal to the difference in number of free parameters in the models. In this case we treat the LS models, with two or three free parameters, as sub-models of standard neutrino mixing with six free parameters, leading to a $\chi^2$-distribution with 4 degrees of freedom when $\eta$ is kept fixed or 3 degrees of freedom when $\eta$ is left as a free parameter. We have verified via Monte-Carlo simulations that the distribution of our $\Delta\chi^2$ test statistic is well approximated by these distributions.

In applying the above formula, the $\chi^2(\Theta)$ is minimised over the parameters $\Theta$ in our fits and is built from three parts;
\begin{equation}
\chi^2(\Theta)=\chi^2_\text{LB}(\Theta)+\chi^2_\text{R}(\Theta)+P(\Theta),
\end{equation}
with $\chi^2_\text{LB}(\Theta)$ for the full simulations of the long-baseline experiments DUNE and T2HK, $\chi^2_\text{R}(\Theta)$ for the constraints from reactor experiments Daya Bay and JUNO, and $P(\Theta)$ for a prior intended to include information from the results of existing experimental measurements. 

For the long-baseline experiments we use the statistical model of the GLoBES
library \cite{Huber:2004ka,Huber:2007ji}, where the $\chi^2_{LB}(\Theta)$ is a
sum of contributions from each of the experiments' channels. The individual
contributions are constructed as
\begin{equation}
\chi^2_c(\Theta)=\min_{\xi=\{\xi_s,\xi_b\}}\left[2\sum_i
\left(\eta_i(\Theta,\xi)-n_i+n_i\ln\frac{n_i}{\eta_i(\Theta,\xi)}\right)+p(\xi,\sigma)\right],
\end{equation}
where $\chi^2_c$ denotes the contribution from a given channel of a given
experiment. The sum in this expression is over the $i$ energy bins of the
experimental configuration, with simulated true event rates of $n_i$ and
simulated event rates $\eta_i(\Theta,\xi)$ for the hypothesis parameters
$\Theta$ and systematic error parameters $\xi$. The systematic errors of the
experiments are treated using the method of pulls, parameterized as $\xi_s$ for
the signal error and $\xi_b$ for the background error. These parameters are
given Gaussian priors which form the term
$p(\xi,\sigma)=\xi_s^2/\sigma_s^2+\xi_b^2/\sigma_b^2$, where
$\sigma=\{\sigma_s,\sigma_b\}$ are the sizes of the systematic errors given by
the experiment.

For the reactor experiments we simply assume independent Gaussian measurements such that
\begin{equation}
\chi^2_\text{R}=\frac{\left(\sin^2{\th13}-\sin^2\overline{\th13}\right)^2}{\sigma_{\th13}^2}+\frac{\left(\sin^2{\th12}-\sin^2\overline{\th12}\right)^2}{\sigma_{\th12}^2}+\frac{\left(\Dm21-\overline{\Dm21}\right)^2}{\sigma_{\Dm21}^2},
\end{equation}
where $\overline{\th13}$, $\overline{\th12}$ and $\overline{\Dm21}$ are the true parameter values and $\sigma_{\th13}$, $\sigma_{\th12}$ and $\sigma_{\Dm21}$ the corresponding experimental measurement uncertainties.

The prior $P(\Theta)$ provides information from existing experimental measurements and is calculated using the results of the NuFIT 3.0 global fit in the same way as our fit in \cref{sec:fit}, so that $P(\Theta)=\chi^2_\text{Fit}(\Theta)$ as defined in \cref{eq:fit}.

In all our simulations, the true parameters are taken to be the best-fit values from the appropriate LS fit results given in \cref{tab:bfp}, except where stated otherwise.

\subsection{Results}\label{sec:res}

\begin{figure}[t]
	\centering
	\sublabel{fig:chisq1d.a}
	\sublabel{fig:chisq1d.b}
	\sublabel{fig:chisq1d.c}
	\sublabel{fig:chisq1d.d}
	\sublabel{fig:chisq1d.e}
	\sublabel{fig:chisq1d.f}
	\includegraphics[width=\textwidth]{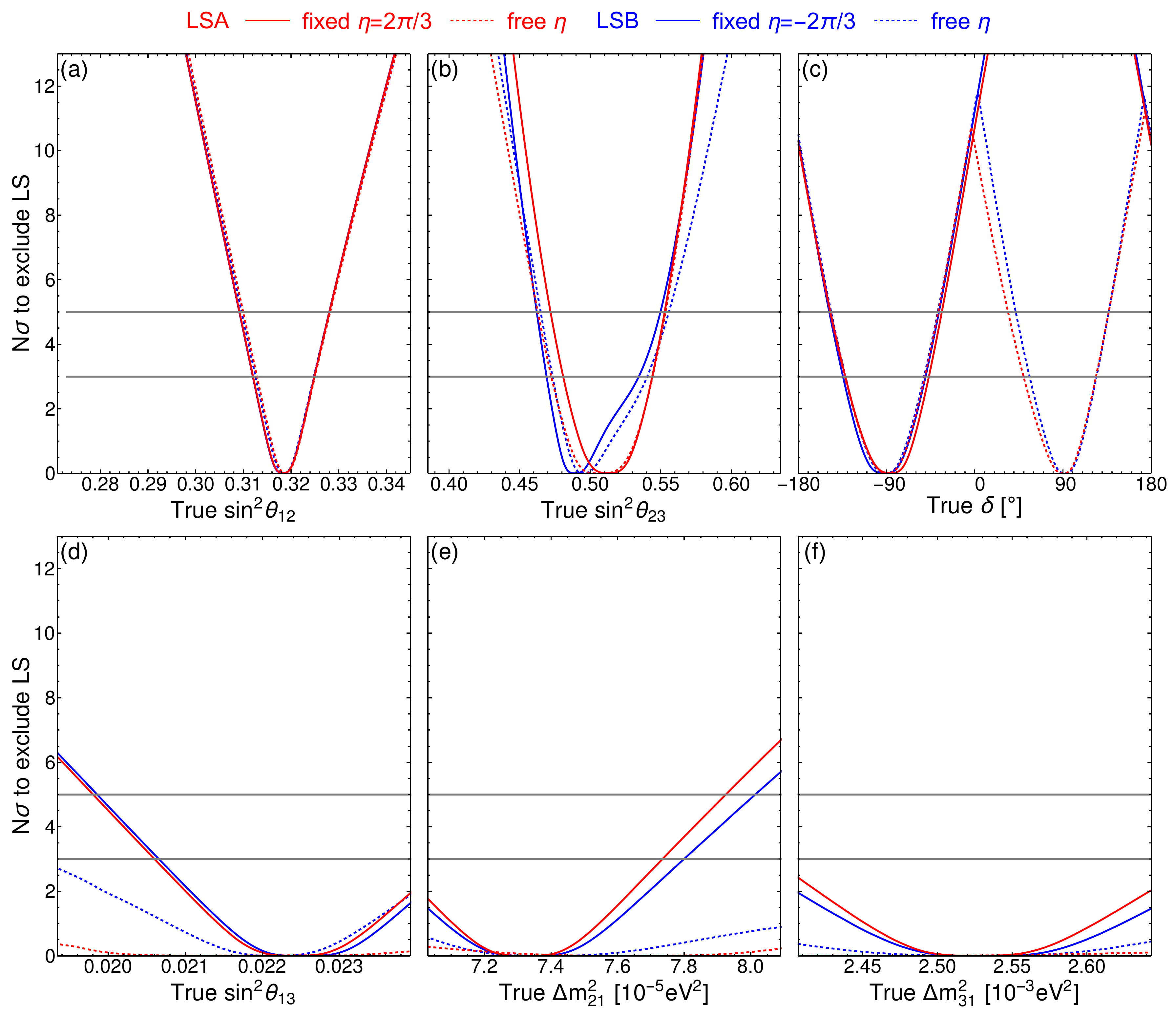}
	\caption{The predicted sensitivity of future experiments to excluding LSA (red) and LSB (blue), shown as a function of the true value of each parameter. Solid curves correspond to the case with $\eta$ fixed at $\eta=\frac{2\pi}{3}$ for LSA or $\eta=-\frac{2\pi}{3}$ for LSB, while dashed curves correspond to the case with $\eta$ left free. The ranges of true parameters shown in the plots
	corresponds to the current three sigma allowed NuFIT 3.0 regions.}
	\label{fig:chisq1d}
\end{figure}

The sensitivity to exclude either version of the LS model is shown as a
function of the true value of each parameter in \cref{fig:chisq1d}, for true
values, with the range selected along the horizontal axes to be that given by the currently allowed at $3\sigma$ by the latest NuFIT 3.0
global fit. In each case, the parameters not shown are assumed to take their
best-fit values from the fit to LS described in \cref{sec:fit}. 

From the upper panels in \cref{fig:chisq1d}, we see that
\th12, \th23 and $\delta$ provide the strongest tests of the model, with there only being 
a relatively small portion of the presently allowed true parameter space where the model would not be
excluded. This is due to the strong predictions of these parameters by the LS
models, as discussed in \cref{sec:pred}. Note that these parameters are those
that will be measured most precisely by the three next-generation experiments
used in our simulations, JUNO, DUNE and T2HK. For these three parameters,
the effect of allowing $\eta$ to vary does not much change the sensitivity,
other than the additional solution (currently disfavoured by experiment) with
$\delta=+90^\circ$ which occurs when changing the sign of $\eta$. For \th12 in
particular there is no effect of allowing $\eta$ to vary. This is due to the
sum rule in \cref{eq:sr1} which relates \th12 with \th13 independently from the
value of $\eta$; the precise measurement of \th13 then fixes the value of \th12
to a narrow range such that a measurement of \th12 outside of this would
exclude the LS model regardless of the LS parameter values. Similarly the
precise measurements of \th13, \Dm21 and \Dm31 strongly constrain the magnitude
(but not sign) of $\eta$, so that the LS allowed regions of the other variables
are not significantly changed when $\eta$ is allowed to vary, with the noted
exception that changing the sign of $\eta$ allows the sign of $\delta$ to also
change.

From the lower panels in \cref{fig:chisq1d}, we see that the sensitivity to exclude LS from measurements of \th13, \Dm21 or \Dm31 is
much less than for the other three parameters and the sensitivity is also
significantly reduced when allowing $\eta$ to vary. By the converse argument to
that used above, this is due to these three parameter measurements driving the
fit to $m_a$ and $m_b$ (and $\eta$), and so a measurement of these parameters
will tend to move the fitted LS parameter values rather than exclude the model,
particularly when fitting the extra free parameter $\eta$. 
%
%
However, a particularly small measurement of \th13 or particularly large
measurement of \Dm21, relative to their current allowed range of values, may
still exclude the fixed $\eta$ version of the models.

\begin{figure}[t]
	\centering
	\sublabel{fig:chisq2d1.a}
	\sublabel{fig:chisq2d1.b}
	\sublabel{fig:chisq2d1.c}
	\sublabel{fig:chisq2d1.d}
	\sublabel{fig:chisq2d1.e}
	\sublabel{fig:chisq2d1.f}
	\includegraphics[width=\textwidth]{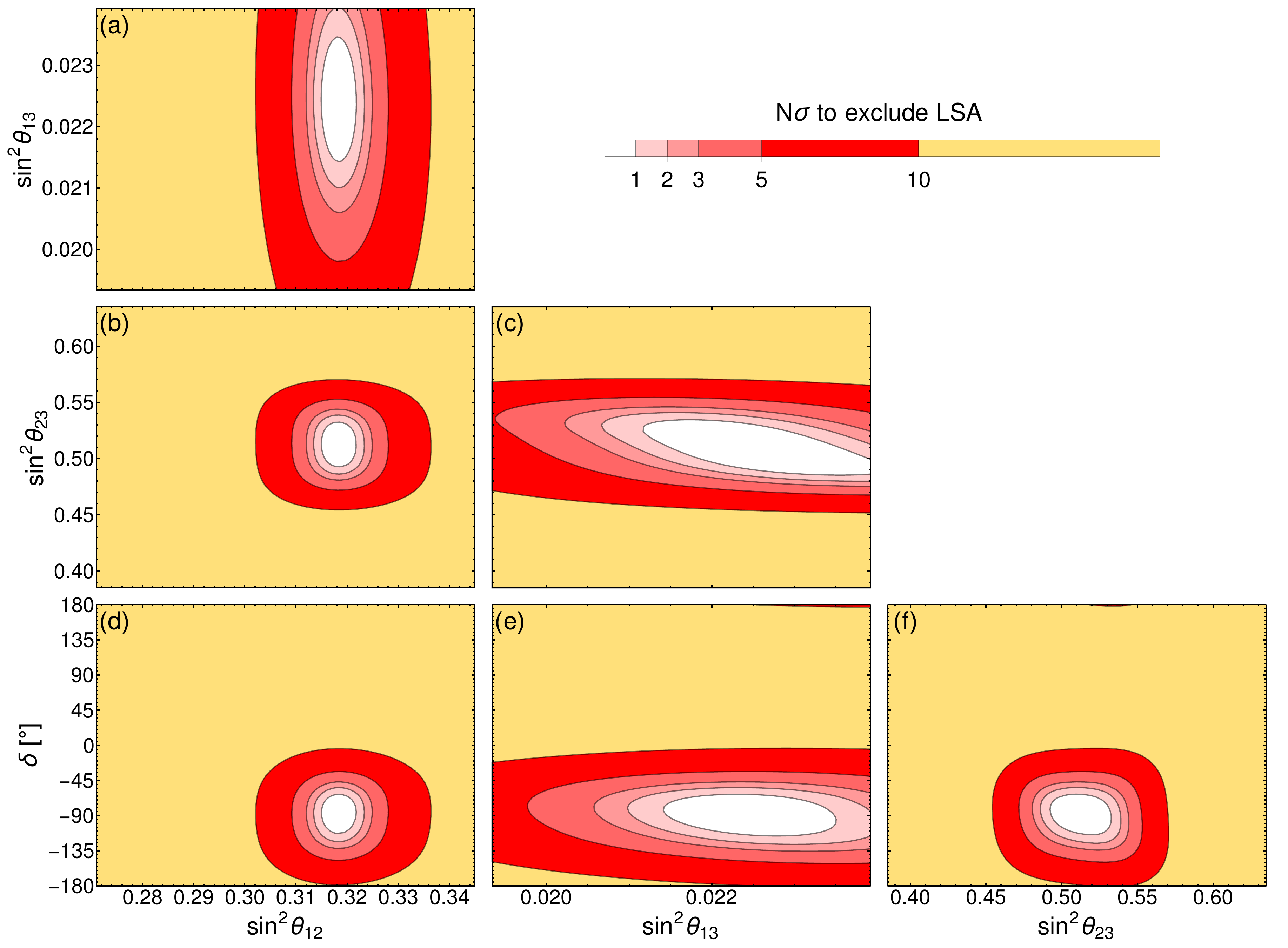}
	\caption{The predicted sensitivity of future experiments to excluding LSA, with $\eta$ fixed at $\eta=\frac{2\pi}{3}$, shown as a function of each pair of true parameters. The ranges of true parameters shown in the plots corresponds to the current three sigma allowed NuFIT 3.0 regions.}
	\label{fig:chisq2d1}
\end{figure}

\begin{figure}[t]
	\centering
	\sublabel{fig:chisq2d1o.a}
	\sublabel{fig:chisq2d1o.b}
	\sublabel{fig:chisq2d1o.c}
	\sublabel{fig:chisq2d1o.d}
	\sublabel{fig:chisq2d1o.e}
	\sublabel{fig:chisq2d1o.f}
	\includegraphics[width=\textwidth]{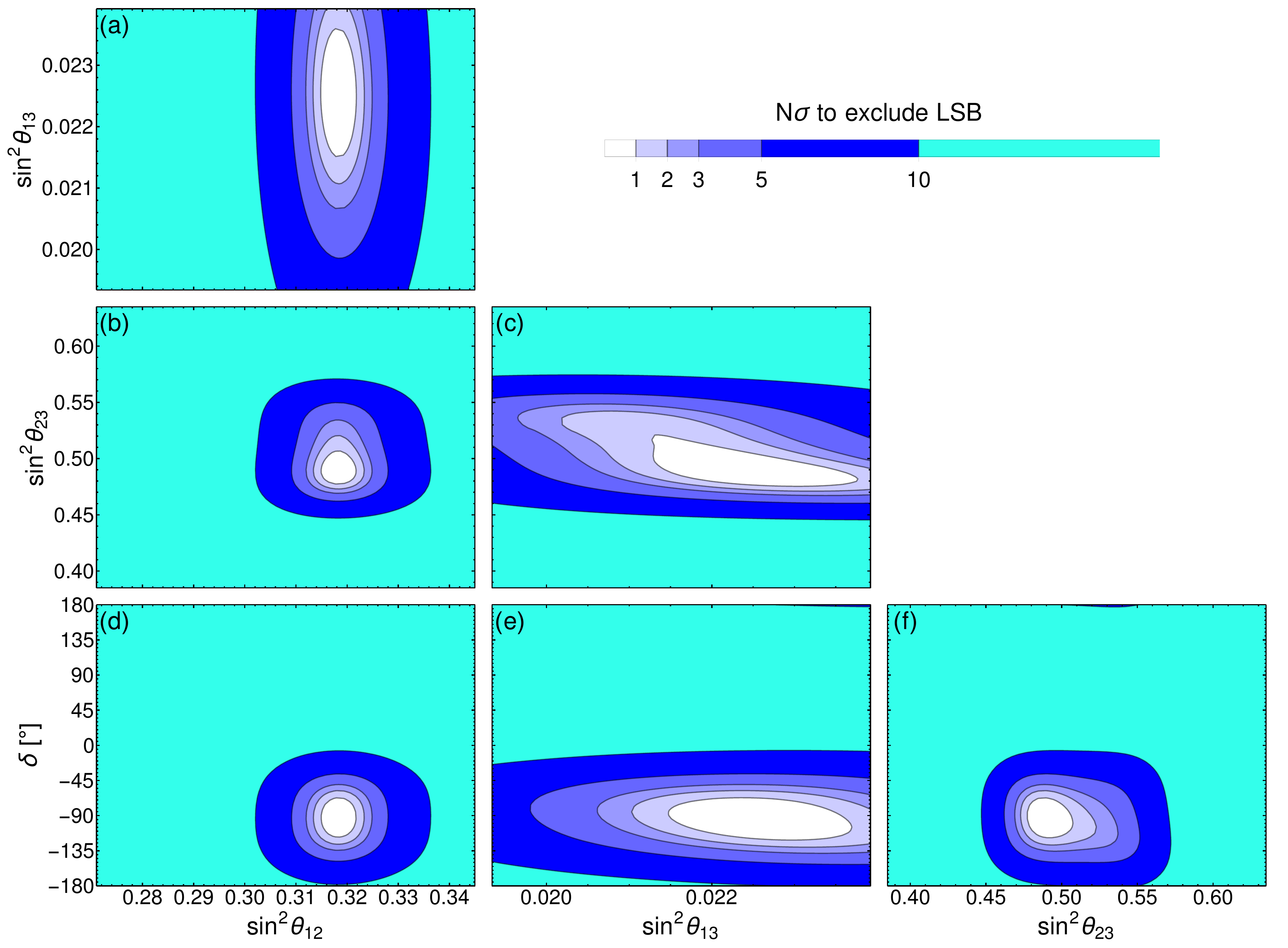}
	\caption{The predicted sensitivity of future experiments to excluding LSB, with $\eta$ fixed at $\eta=-\frac{2\pi}{3}$, shown as a function of each pair of true parameters. The ranges of true parameters shown in the plots corresponds to the current three sigma allowed NuFIT 3.0 regions.}
	\label{fig:chisq2d1o}
\end{figure}

\begin{figure}[ht]
	\centering
	\sublabel{fig:chisq2d2.a}
	\sublabel{fig:chisq2d2.b}
	\sublabel{fig:chisq2d2.c}
	\sublabel{fig:chisq2d2.d}
	\sublabel{fig:chisq2d2.e}
	\sublabel{fig:chisq2d2.f}
	\sublabel{fig:chisq2d2.g}
	\sublabel{fig:chisq2d2.h}
	\sublabel{fig:chisq2d2.i}
	\includegraphics[width=\textwidth]{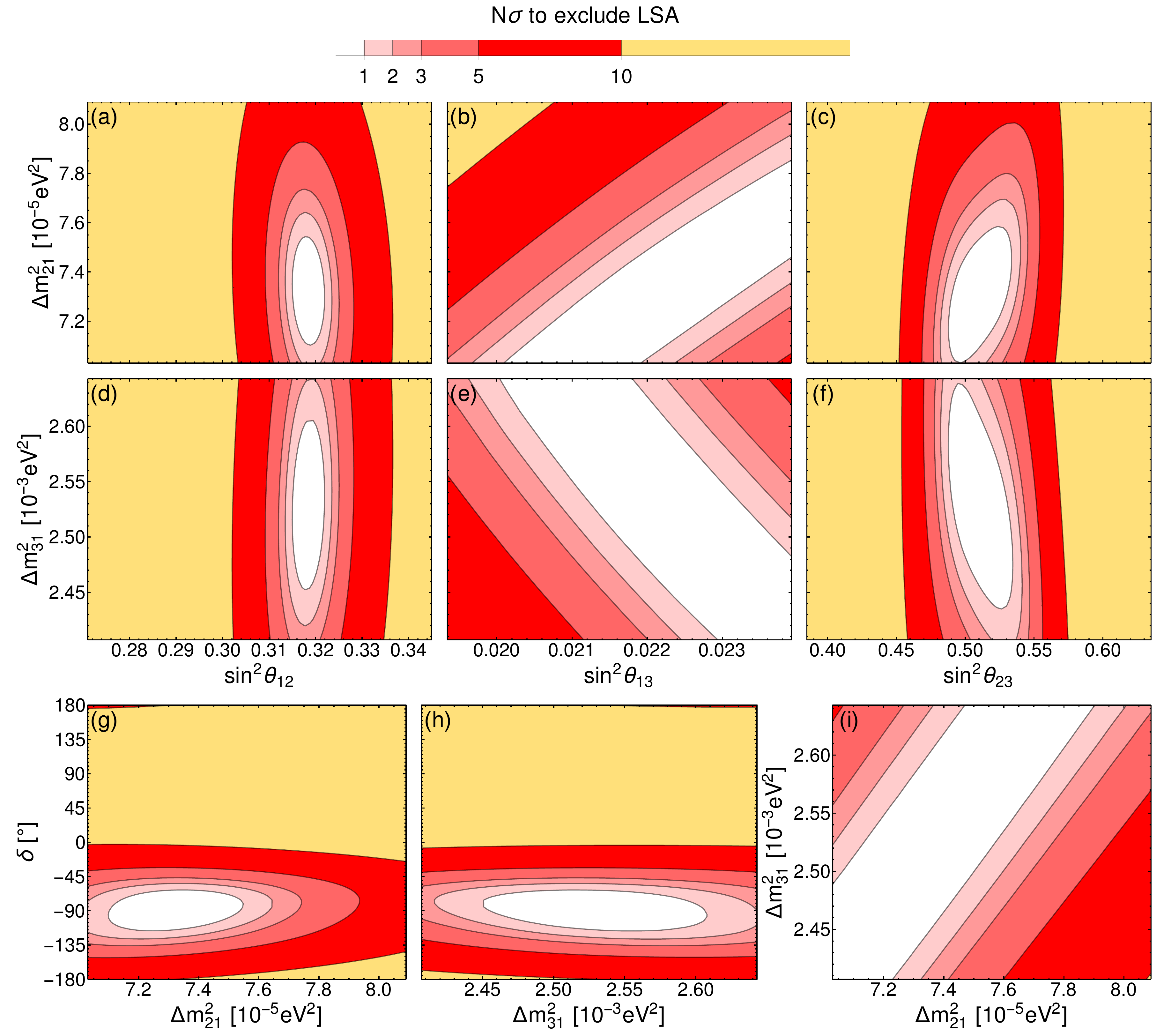}
	\caption{The predicted sensitivity of future experiments to excluding LSA, with $\eta$ fixed at $\eta=\frac{2\pi}{3}$, shown as a function of each pair of true parameters. The ranges of true parameters shown in the plots corresponds to the current three sigma allowed NuFIT 3.0 regions.}
	\label{fig:chisq2d2}
\end{figure}

\begin{figure}[ht]
	\centering
	\sublabel{fig:chisq2d2o.a}
	\sublabel{fig:chisq2d2o.b}
	\sublabel{fig:chisq2d2o.c}
	\sublabel{fig:chisq2d2o.d}
	\sublabel{fig:chisq2d2o.e}
	\sublabel{fig:chisq2d2o.f}
	\sublabel{fig:chisq2d2o.g}
	\sublabel{fig:chisq2d2o.h}
	\sublabel{fig:chisq2d2o.i}
	\includegraphics[width=\textwidth]{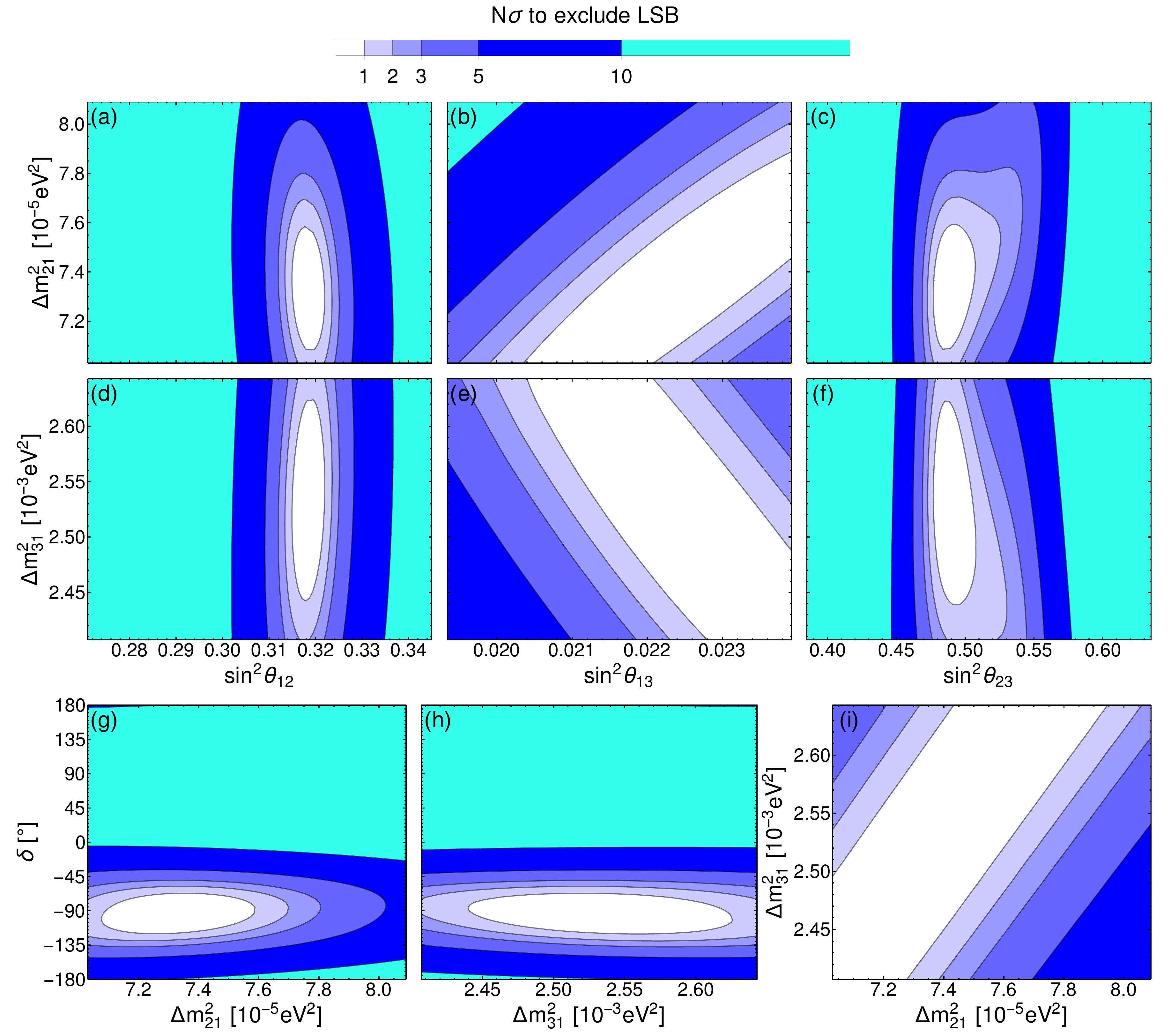}
	\caption{The predicted sensitivity of future experiments to excluding LSB, with $\eta$ fixed at $\eta=-\frac{2\pi}{3}$, shown as a function of each pair of true parameters. The ranges of true parameters shown in the plots corresponds to the current three sigma allowed NuFIT 3.0 regions.}
	\label{fig:chisq2d2o}
\end{figure}

The results shown in \cref{fig:chisq1d} show only the dependence of the
significance to exclude LS on the true value of each variable
individually.  However, the sensitivity will generally have a
strong dependence on the true values of the other parameters.
The significance to exclude the LS models depending on the true values of each
pair of variables, for the cases where $\eta$ is kept fixed, is shown in
\Cref{fig:chisq2d1,fig:chisq2d2} for LSA and in
\cref{fig:chisq2d1o,fig:chisq2d2o} for LSB. 

Each panel of 
\cref{fig:chisq2d1,fig:chisq2d1o} includes two dimensionless variables 
(i.e. angle or phase) which both depend only
on the ratio of LS input parameters $r=m_b/m_a$, and so, in a LS model, a measurement of any one of these parameters corresponds
to a measurement of $r=m_b/m_a$ (see Fig.~\ref{fig:params-ratio}). Combining two of these parameter measurement therefore
give two measurements of $r=m_b/m_a$, with any conflict between them providing strong
evidence to exclude the model. For this reason the significance to exclude the
models is close to being simply the combined significance from individual
measurements implied by \cref{fig:chisq1d}. 

By contrast, each panel of \Cref{fig:chisq2d2,fig:chisq2d2o}
shows the results for the pairs of variables including at least one dimensionful mass-squared
difference. Here we can see in
\cref{fig:chisq2d2.b,fig:chisq2d2.e,fig:chisq2d2.i} for LSA, and in
\cref{fig:chisq2d2o.b,fig:chisq2d2o.e,fig:chisq2d2o.i} for LSB, there is a
strong correlation between the measurements of \th13, \Dm21 and \Dm31. This shows
clearly that, although individual measurements of these parameters cannot exclude a LS model
(since the parameters of the LS model could be adjusted to accommodate any of them individually) 
a {\em combined} measurement of two of them could serve to exclude the model.
This is the reason for presenting these combined sensitivity plots.
Of the three parameters for which such 
{\em combined} measurements provide the strongest test of the model, each pair
includes measurements from different experiments, with \th13 coming mainly from
the short-baseline reactor measurement such as Daya Bay, \Dm21 from the medium-baseline reactor
measurement such as JUNO, and \Dm31 from the long-baseline accelerator measurement such as DUNE and T2HK. This
demonstrates a strong synergy between all these experiments in attempts to exclude the LS
models.



\section{\label{sec:conc}Conclusion}

In this paper, we have investigated the ability to probe
one of the most predictive viable neutrino mass and mixing models
with future neutrino oscillation experiments: the Littlest Seesaw.  The LS
models work within the framework of the Type I
seesaw mechanism, using two right-handed neutrinos to generate the left-handed
neutrino masses. Combined with constraints from flavour symmetries, the
neutrino mixing angles and phases can be predicted from a small number of
parameters; in its most constrained form all neutrino masses, angles, and
phases are determined from just two input parameters. In fact, we have shown
that while the neutrino masses depend on the two mass parameters
independently, the mixing angles and phases 
depend only on a single dimensionless quantity, the ratio of these two input 
parameters.

We have studied two versions of this model (LSA and LSB) which use different flavour
symmetries to enforce constraints which result in different permutations of the
second and third rows and columns of the neutrino mass matrix, leading to
different predictions for the octant of \th23. Using the results of a recent
global fit of neutrino oscillation experiments, we have found that both
versions can well accommodate the parameter values as measured by
experiment, with the greatest tension on the value of \th23
at the $1\sigma$ level. The prediction of LS is very close to the maximal
mixing value with experimental results from NO$\nu$A suggesting a more
non-maximal value, while results from T2K still consistent with a maximal value
of \th23. We find that the LSB version, predicting a value of
\th23 in the lower octant, to be slightly preferred.

The ability of future experiments to exclude these models then comes from a
convolution of the strength of the predictions of the model with the
sensitivity of the experiments in measuring those parameters. Through our fit
of the models to current global neutrino oscillation data, we have seen that the LS models
make strong predictions for the values of \th12, \th23, and $\delta$, the three
parameters for which current measurements are weakest. In addition we find that,
for certain combinations of the remaining observables, \th13, \Dm21 and \Dm31, the LS models predict strong
correlations. 

With future experiments expected to improve precision on all six
parameters measured through oscillations, our simulations have shown that the LS
models can be thoroughly tested through future precise individual measurements of \th12,
\th23, and $\delta$.
This can be readily understood since the free parameters of the LS models are currently most 
constrained by the precise measurements of \th13, \Dm21 and \Dm31,
leading to predictions for the currently less well determined parameters \th12, \th23, and $\delta$. 

The predictivity of the LS models means that an even higher precision measurement of those parameters which currently drive the fit
of the input parameters, namely \th13, \Dm21
and \Dm31, could still exclude the LS models when considered in combination with each other.
For example, the combination of any
two of them could require a region of LS parameter space already excluded by the third.

These above results all highlight the strong complementarity between different classes of
oscillation experiment. While the long baseline accelerator experiments DUNE
and T2HK are expected to provide the strongest measurements of \th23 and
$\delta$ (two of those that can {\em individually} test the model's viability)
the third, \th12, will come from medium baseline reactor experiments such as
JUNO and RENO-50. The strongest complementarity, however, comes
from combining precision measurements of $\Dm21$,
$\Dm31$ and $\th13$, where any pair of these measurements relies
on the results from all the different experiments: long-baseline accelerator
experiments for \Dm31, medium-baseline reactor experiments for \Dm21, and
short-baseline reactor experiments for \th13.

In summary, the work presented in this paper shows that the most straightforward way to exclude the LS model
is to provide a better {\em individual} determination of 
the three currently less precisely measured parameters \th12, \th23, and $\delta$,
which requires both medium baseline experiments such as JUNO and RENO-50, and long baseline experiments such as DUNE and T2HK,
where the synergy between the latter two experiments is thoroughly explored in \cite{synergy}.
In addition, the LS model could be constrained by {\em combined} measurements of the three
remaining parameters $\Dm21$, $\Dm31$ and $\th13$, where an even higher precision of the latter reactor parameter
at the short baseline Daya Bay experiment can also play an important role.

We remark that, although the above conclusions have been established
for the LSA and LSB models, similar arguments can be
expected to apply to any highly predictive flavour models which determine the
oscillation parameters from a smaller number of input model parameters. In any
such model, the input parameters will tend to be tuned to fit the strong
constraints from the most precisely measured parameters, leading to predictions
of the other parameters. If the models can accommodate individual measurements in this way,
distinguishing between them using those parameters which drive the fit is still
possible, if those models are highly constrained, but this requires the
parameter measurements to be considered in combination. 

In conclusion, the need
for future reactor and accelerator experiments to measure individually \th12, \th23 and
$\delta$, plus combinations of \th13, \Dm21 and \Dm31, may be considered 
to be general requirements in order to probe predictive flavour symmetry models.
Therefore a broad programme of such precision experiments seems to be 
essential in order to take the next step in understanding neutrino oscillations in the
context of the flavour puzzle of the Standard Model.

\acknowledgments

We would like to thank Michel Sorel, Alan Bross and Ao Liu for providing
experimental information for use in our simulation of DUNE, and also
the Hyper-Kamiokande proto-collaboration collaboration for information used in
our simulations for Hyper-Kamiokande.

PB, SP and TC acknowledge partial support from the European Research Council
under ERC Grant ``NuMass'' (FP7-IDEAS-ERC ERC-CG 617143).
We all acknowledge partial support from
ELUSIVES ITN (H2020-MSCA-ITN-2015, GA-2015-674896-ELUSIVES), and InvisiblesPlus RISE
(H2020-MSCARISE-2015, GA-2015-690575-InvisiblesPlus).
SP gratefully acknowledges partial support from the Wolfson Foundation and the Royal Society.

\appendix
\section*{Appendix}
\section{Exact expressions for LS sum rules}\label{app:sr}
The angles and Dirac phase can then be written as
\begin{equation}\label{eq:params-ratio}
\sin^2\th13=s(r),\qquad\tan^2\th12=t(r),\qquad\cos2\theta_{23}=\pm
c(r),\qquad\cos\delta=\pm d(r), \end{equation}
with positive signs taken for LSA and negative for LSB and where
\begin{align}
s(r)=&\frac{1}{6}\left(1-\frac{55r^2+4(1-4r)}{\sqrt{\left((11r)^2+4(1-7r)\right)\left((11r)^2+4(1-r)\right)}}\right)\\
t(r)=&\frac{1}{4}\left(1+\frac{55r^2+4(1-4r)}{\sqrt{\left((11r)^2+4(1-7r)\right)\left((11r)^2+4(1-r)\right)}}\right)\\
c(r)=&\frac{2r(11r-1)\left(55r^2-16r+4-5\sqrt{\left((11r)^2+4(1-7r)\right)\left((11r)^2+4(1-r)\right)}\right)}{\left((11r)^2+4(1-7r)\right)\left((11r)^2+4(1-r)\right)+4r^2\left((11r)^2+2(2-11r)\right)}\\
d(r)=&-\frac{c(r)(1-5s(r))}{2\sqrt{2s(r)(1-c(r)^2)(1-3s(r))}}.
\end{align}
Similar expressions for
the Majorana phases also possible. Combining these, expressions relating any
two of the angles and/or phases can be found. The first such relation, relating
\th13 and \th12, is the same as \cref{eq:sr1}, which is general for all
CSD($n$). New exact relations between \th13 and \th23 or \th12 and \th23, as
well as the relation between $\delta$ and \th12, true for LSA with
$\eta=\frac{2\pi}{3}$ or LSB with $\eta=-\frac{2\pi}{3}$, are found of the form
\begin{equation}
f_\pm(\th13,\th23)=0,\qquad g_\pm(\th12,\th23)=0,\qquad h_\pm(\delta,\th12)=0,\label{eq:sr}
\end{equation}
where again the positive (negative) sign is used in the functions valid for LSA
(LSB). Exact expressions are given as
\begin{align}
f_\pm&(\th13,\th23)=\frac{44\s13^2\sqrt{1-3\s13^2}}{4(1-6\s13^2)\mp 3\c13^2\cos2\th23}\pm\frac{\c13^2\cos2\th23}{\sqrt{1-3\s13^2}}-\sqrt{\frac{8\s13^2}{3}-\frac{\c13^4\cos^22\th23}{3(1-3\s13^2)}},\\
g_\pm&(\th12,\th23)=\frac{22\s12^2\sqrt{1-3\s12^2}}{2(5\s12^2-1)\mp\cos2\th23}\pm\frac{\cos2\th23}{\sqrt{1-3\s12^2}}-\sqrt{4\s12^2-\frac{\cos^22\th23}{3(1-3\s12^2)}},\\
h_\pm&(\delta,\th12)=\frac{5\s12^2-1}{\s12\sqrt{1-3\s12^2}}\pm\frac{\sqrt{3}\cos\delta}{\sqrt{1-12\s12^2(1-3\s12^2)\sin^2\delta}}+\frac{11\sqrt{1-12\s12^2(1-3\s12^2)\sin^2\delta}}{2(6\s12^2-1)\sin\delta\mp2\sqrt{3}\cos\delta}.
\end{align}

\end{document}